\documentclass[12pt]{article}
\usepackage{epsf,latexsym}
\usepackage{amsmath,amsfonts,amssymb,amsthm,cite}

\newtheorem{thm}{Theorem}[section]
\newtheorem{lem}[thm]{Lemma}
\newtheorem{cly}[thm]{Corollary}

\newtheorem{conj}[thm]{Conjecture}

\theoremstyle{definition}                    
\newtheorem{defn}[thm]{Definition}
\theoremstyle{remark}
\newtheorem{rem}[thm]{Remark}

\numberwithin{equation}{section}             

\usepackage[ps,arc,frame]{xy}

\newcommand{\rxy}[1]{{\begin{xy}
0;<2mm,0mm>:<0mm,2mm>::0;0,#1\end{xy}}}

\newcommand{\rxyz}[2]{{\begin{xy} 0;<2mm,0mm>:<0mm,2mm>::0;0,
,(5,-2)*{a} ,(10,-2)*{b} ,(15,-2)*{c} ,(2,-5)*{a} ,(2,-10)*{b} ,(2,-15)*{c}
,(5,-5)*\cir(#1,0){} ,(10,-5)*\cir(#1,0){} ,(15,-5)*\cir(#1,0){}
,(5,-10)*\cir(#1,0){} ,(10,-10)*\cir(#1,0){} ,(15,-10)*\cir(#1,0){}
,(5,-15)*\cir(#1,0){} ,(10,-15)*\cir(#1,0){} ,(15,-15)*\cir(#1,0){}
#2\end{xy}}}

\newcommand{\rxyzz}[2]{{\begin{xy} 0;<2mm,0mm>:<0mm,2mm>::0;0,
,(5,-2)*{a} ,(10,-2)*{b} ,(15,-1.8)*{\bar{b}} ,(2,-5)*{a} ,(2,-10)*{b}
,(2,-14.8)*{\bar{b}} ,(5,-5)*\cir(#1,0){} ,(10,-5)*\cir(#1,0){} ,(15,-5)*\cir(#1,0){}
,(5,-10)*\cir(#1,0){} ,(10,-10)*\cir(#1,0){} ,(15,-10)*\cir(#1,0){}
,(5,-15)*\cir(#1,0){} ,(10,-15)*\cir(#1,0){} ,(15,-15)*\cir(#1,0){}
#2\end{xy}}}

\newcommand{\rxyb}[2]{{\begin{xy} 0;<2mm,0mm>:<0mm,2mm>::0;0,
,(5,-2)*{a} ,(10,-1.8)*{\bar{a}} ,(15,-2)*{b} ,(20,-1.8)*{\bar{b}} ,(2,-5)*{a}
,(1.8,-10)*{\bar{a}} ,(2,-15)*{b} ,(1.8,-20)*{\bar{b}} ,(5,-5)*\cir(#1,0){}
,(10,-5)*\cir(#1,0){} ,(15,-5)*\cir(#1,0){} ,(20,-5)*\cir(#1,0){}
,(5,-10)*\cir(#1,0){} ,(10,-10)*\cir(#1,0){} ,(15,-10)*\cir(#1,0){}
,(20,-10)*\cir(#1,0){} ,(5,-15)*\cir(#1,0){} ,(10,-15)*\cir(#1,0){}
,(15,-15)*\cir(#1,0){} ,(20,-15)*\cir(#1,0){} ,(5,-20)*\cir(#1,0){}
,(10,-20)*\cir(#1,0){} ,(15,-20)*\cir(#1,0){} ,(20,-20)*\cir(#1,0){}
#2\end{xy}}}

\newcommand{\rxyc}[2]{{\begin{xy} 0;<2mm,0mm>:<0mm,2mm>::0;0,
,(5,-2)*{a} ,(10,-1.8)*{\bar{a}} ,(2,-5)*{a} ,(1.8,-10)*{\bar{a}}
,(5,-5)*\cir(#1,0){} ,(10,-5)*\cir(#1,0){} ,(5,-10)*\cir(#1,0){}
,(10,-10)*\cir(#1,0){} #2\end{xy}}}

\newcommand{\rxycc}[2]{{\begin{xy} 0;<2mm,0mm>:<0mm,2mm>::0;0,
,(5,-2)*{\mu_{ij}} ,(10,-2)*{\mu_{ik}} ,(2,-5)*{\mu_{ij}}
,(2,-10)*{\mu_{\ell j}} ,(5,-5)*\cir(#1,0){} ,(10,-5)*\cir(#1,0){}
,(5,-10)*\cir(#1,0){} ,(10,-10)*\cir(#1,0){} #2\end{xy}}}

\newcommand{\rxyd}[2]{{\begin{xy} 0;<2mm,0mm>:<0mm,2mm>::0;0,
,(5,-2)*{a} ,(10,-1.8)*{\bar{a}} ,(15,-2)*{b} ,(20,-1.8)*{\bar{b}} ,(25,-2)*{c}
,(2,-5)*{a} ,(2,-10)*{\bar{a}} ,(2,-15)*{b} ,(2,-20)*{\bar{b}} ,(2,-25)*{c}
,(5,-5)*\cir(#1,0){} ,(10,-5)*\cir(#1,0){} ,(15,-5)*\cir(#1,0){}
,(20,-5)*\cir(#1,0){} ,(25,-5)*\cir(#1,0){} ,(5,-10)*\cir(#1,0){}
,(10,-10)*\cir(#1,0){} ,(15,-10)*\cir(#1,0){} ,(20,-10)*\cir(#1,0){}
,(25,-10)*\cir(#1,0){} ,(5,-15)*\cir(#1,0){} ,(10,-15)*\cir(#1,0){}
,(15,-15)*\cir(#1,0){} ,(20,-15)*\cir(#1,0){} ,(25,-15)*\cir(#1,0){}
,(5,-20)*\cir(#1,0){} ,(10,-20)*\cir(#1,0){} ,(15,-20)*\cir(#1,0){}
,(20,-20)*\cir(#1,0){} ,(25,-20)*\cir(#1,0){} ,(5,-25)*\cir(#1,0){}
,(10,-25)*\cir(#1,0){} ,(15,-25)*\cir(#1,0){} ,(20,-25)*\cir(#1,0){}
,(25,-25)*\cir(#1,0){} #2\end{xy}}}

\epsfverbosetrue
\newcommand{\double}[1]{\mathbb{#1}}

\newcommand{\cc}{\double{C}}
\newcommand{\nn}{\double{N}}
\newcommand{\rr}{\double{R}}
\newcommand{\zz}{\double{Z}}

\newcommand{\kk}{\double{K}}

\newcommand{\Aa}{\mathcal{A}}
\newcommand{\ccc}{\mathcal{C}}

\newcommand{\hhh}{\double{H}}
\newcommand{\mm}{\mathcal{M}}

\newcommand{\dd}{\mathcal{D}}
\newcommand{\mmf}{\hbox{$^f$\hspace{-0.2cm} $\mathcal{M}$}}            
\newcommand{\Mf}{\hbox{$^f$\hspace{-0.2cm} {\it M}}}                    
\newcommand{\MfT}{\hbox{$^{f^T}$\hspace{-0.19cm} {\it M}}}               
\newcommand{\Mfm}{\hbox{$^{\hat{f}}$\hspace{-0.2cm} {\it M}}}
\newcommand{\Mg}{\hbox{$^g$\hspace{-0.2cm} {\it M}}}
\newcommand{\Mgf}{\hbox{$^{gf}$\hspace{-0.2cm} {\it M}}}          
\newcommand{\ddf}{\hbox{$^f$\hspace{-0.15cm} $\mathcal{D}$}}           
\newcommand{\ddfm}{\hbox{$^{\hat{f}}$\hspace{-0.15cm}
$\mathcal{D}$}}   
\newcommand{\hh}{\mathcal{H}}

\newcommand{\LLL}{\mathcal{L}}
\newcommand{\sss}{\mathcal{S}}

\newcommand{\T}{{\rm tr}}

\newcommand{\ot}{\otimes}
\newcommand{\op}{\oplus}

\newcommand{\bb}{\begin{eqnarray}}
\newcommand{\ee}{\end{eqnarray}}
\newcommand{\eee}{\nonumber\end{eqnarray}}
\newcommand{\pp}[1]{\begin{pmatrix} #1 \end{pmatrix}}
\newcommand{\qq}{\quad}

\begin{document}
\font\twelve=cmbx10 at 13pt
\font\eightrm=cmr8

\thispagestyle{empty}

\begin{center}

CENTRE DE PHYSIQUE TH\'EORIQUE \footnote{\, Unit\'e Propre de
Recherche 7061} \\ CNRS--Luminy, Case 907\\ 13288 Marseille Cedex 9\\
FRANCE\\

\vspace{2cm}

{\Large\textbf{On a Classification of Irreducible \\ Almost Commutative
Geometries}} \\

\vspace{1.5cm}

{\large Bruno Iochum
\footnote{\, and Universit\'e de Provence,\\
${}$\qq\qq\qq\  iochum@cpt.univ-mrs.fr \qq
\qq schucker@cpt.univ-mrs.fr \qq\qq}, Thomas Sch\"ucker$\,\,^2$,
Christoph Stephan
\footnote{\, and Universit\'e de Provence and Universit\"at Kiel,\\
${}$\qq\qq\qq\ stephan@cpt.univ-mrs.fr} }

\vspace{1.5cm}

{\large\textbf{Abstract}}
\end{center}

We classify all irreducible, almost commutative geometries whose spectral 
action is dynamically non-degenerate. Heavy use is made of Krajewski's 
diagrammatic language. 
The motivation for our definition of dynamical
non-degeneracy stems from particle physics where the fermion masses 
are non-degenerate.

\vspace{1.5cm}

\vskip 1truecm

PACS-92: 11.15 Gauge field theories\\
\indent MSC-91: 81T13 Yang-Mills and other gauge theories

\vskip 1truecm

\noindent December 2003
\vskip 1truecm
\noindent CPT-2003/P.4605\\
\noindent hep-th/yymmxxx

\newpage
\tableofcontents
\newpage

\section{Introduction}

Within noncommutative geometry pioneered by Connes 
\cite{book,real,grav,cc},  the almost commutative ones play an interesting role.
They  are defined by spectral triples ($\Aa,\hh,\dd$) where the 
algebra $\Aa_t$ has the form $\Aa_t=C^{\infty}(M) \otimes \Aa$ 
with $\Aa$  a direct sum of  matrix algebras 
and $M$ a (compact Euclidean) spacetime. For instance in 
the standard model of particle physics, 
$\Aa= \hhh \oplus\cc \oplus  M_3(\cc)$. 
It is important to classify the almost commutative triples because of their
applications to physics. Let us sketch some of them.

Einstein's derivation of general relativity from
Riemannian geometry goes in two steps. The first step sets
up the kinematics: the equivalence principle uses general
coordinate transformations and starts from the flat metric of
special relativity to guess curved metrics. The second step
constructs a dynamics for the set of all metrics by
imposing covariance under general coordinate
transformations. 

Connes generalizes Einstein's derivation
to noncommutative geometry \cite{grav,cc}. In this new
setting the metric is encoded in a Dirac operator and a
coherent definition of the equivalence principle becomes
available: the fluctuations of the Dirac operator by algebra
automorphisms properly lifted to the Hilbert space of
spinors. Indeed in Riemannian geometry, the algebra is the
commutative algebra of functions on spacetime, the
automorphisms are precisely the general coordinate
transformations and fluctuating the flat Dirac operator
leads to Dirac operators with curvature and
torsion. The second step is the spectral action which
in the commutative i.e.
Riemannian case reproduces the Einstein-Hilbert action plus a positive
cosmological constant and a curvature squared term. The Euclidean
spectral  action is positive definite and its ground states can be
interpreted as a regularization of the initial singularity.

A noncommutative space or geometry is defined by a `spectral triple'
consisting of an algebra $\Aa$, a Hilbert space $\hh$ and a Dirac operator $\dd$.
One important property of noncommutative geometry is that it contains
discrete spaces, commutative or not. They have finite dimensional
algebras and Hilbert spaces. An almost commutative geometry is a
tensor product of the infinite dimensional commutative algebra
consisting of spacetime functions with a finite dimensional
noncommutative algebra, the `internal space'. The internal Dirac
operator is simply an initial fermionic mass matrix and the internal
algebra automorphisms of the tensor product are gauge
transformations. The internal fluctuations produce gauge bosons and
Higgs scalars and the spectral action of an almost commutative
geometry produces -- besides the gravitational action -- the complete
Yang-Mills-Higgs action. The Higgs scalar is therefore the internal
metric and its dynamics is the Higgs potential. The initial fermionic
mass matrix in general is not a solution of the internal dynamics, the
`internal Einstein equation'. These solutions are the minima of the
Higgs potential which induces the spontaneous symmetry breaking.
They yield the true fermionic mass matrix and we have to compute it.
Let us remark that in the prior approach of noncommutative 
Yang-Mills theories without gravity \cite{cl}, the Dirac operator was
not
 a dynamical variable. Consequently there was no nuance between 
initial and true fermion masses.

Although only a small subset of all Yang-Mills-Higgs models can be
described as an almost commutative geometry this subset is still
infinite and difficult to assess. We propose to reduce it using two
constraints. The first is inspired by `grand unified theories', in
particular 
$SO(10)$: the gauge group of the standard model of electro-magnetic,
weak and strong forces is embedded into a simple group and the
representation of one generation of quarks and leptons is embedded
into an irreducible representation. Therefore our first constraint is:
take the internal algebra simple and its spectral triple
irreducible. As we will see the resulting fermion masses in the
ground state are degenerate in flagrant contradiction to experiment.
We therefore analyze internal algebras with two and three simple
summands and their irreducible spectral triples. Again, in most cases
the fermion masses come out degenerate with a few exceptions. Our
aim is to list these exceptions. In other words, our second constraint is 
to impose a
non-degenerate fermionic mass spectrum in the ground state, namely 
to restrict the analysis to dynamically non-degenerate spectral 
triples. The mathematical definitions of these constraints are given 
in sections 2  and 4.

\section{Irreducibility}

A spectral triple is given by  ($\Aa,\hh,\dd$)  such that the real *-algebra
$\Aa$ acts on the complex Hilbert space $\hh$,  
the Dirac operator $\dd$ on $\hh$ is
selfadjoint and a priori unbounded. These three items
 satisfy certain constraints of geometrical significance
\cite{book,real,grav}. The commutative examples come from the triple
($\Aa=\ccc^{\infty}(M),\hh= \LLL^2 (M,\sss),
\dd=i\gamma^{\mu}\partial_{\mu})$ associated to a compact Riemannian
spin manifold $M$. As the resolvent of $\dd$ is compact,
$(\Aa,\hh,\dd=0)$ is never a spectral triple for infinite dimensional
$\hh$. However, this degenerate situation can occur in finite cases, but is
excluded from our definition of irreducible spectral triples.

\begin{defn}
 i) A spectral triple $(\Aa,\hh,\dd)$ is {\it degenerate} if the kernel of
$\dd$ contains a non-trivial subspace of the complex Hilbert space $\hh$
invariant under the representation $\rho$ on $\hh$ of the real algebra
 $\Aa$.  \\
 ii) A non-degenerate spectral triple $(\Aa,\hh,\dd)$ is {\it reducible} if
there is a proper subspace
$\hh_0\subset\hh$ invariant under the algebra $\rho(\Aa)$  such that
$(\Aa,\hh_0,\dd|_{\hh_0})$ is a non-degenerate spectral triple. If the
triple is real, $S^0$-real and even, we require  the subspace
$\hh_0$ to be also invariant under the real structure $J$, the $ S^0$-real
structure $\epsilon $ and under the chirality
$\chi $ such that the triple $(\Aa,\hh_0,\dd|_{\hh_0})$ is again real,
$S^0$-real and even.
\end{defn}

\begin{rem}
 i) ($\Aa=\ccc^{\infty}(M),\hh= \LLL^2(M,\sss),
\dd=i\gamma^{\mu}\partial_{\mu})$ is never degenerate.\\
\indent ii) If $(\Aa_i,\hh_i,\dd_i)$ are two spectral triples then 
$(\Aa_1 \otimes\Aa_2,\hh_1\otimes \hh_2,\dd_1\otimes 1)$ is a spectral
triple whose kernel of $\dd_1\otimes 1$ is infinite dimensional when 
$\hh_2$ is of infinite dimension.\\
 \indent iii) A finite dimensional commutative triple is a collection of
points. It is non-degenerate if all points have finite distances.
 The converse is wrong, $\Aa=\cc\op\cc\op\cc$ with the triple given by
diagram 15 is a counterexample.\\
 \indent iv) A reducible triple is not necessarily decomposable into a direct
sum. For example,
\bb\Aa=\cc\op\cc\owns (a,b),\qq 
\rho (a,b)=\pp{\pp{a&0\\ 0&\bar a}&0&0&0\\0&b&0&0\\ 0&0&\pp {\bar
a&0\\ 0&\bar a}&0\\0&0&0&\bar a},\ee
\bb \dd=\pp{0&\mm&0&0\\
\mm^*&0&0&0\\ 0&0&0&\overline{\mm}\\
0&0&\overline{\mm^*}&0},\qq\mm=\pp{1\\2}.\ee
Here and throughout, $\bar{ }$ and $^*$ mean complex conjugation and 
adjoint.\\
 \indent v) Our definition of irreducibility differs from the one in our
favorite book \cite{costa}. Indeed every spectral triple coming from a
Riemannian spin manifold is irreducible in our case, even if the manifold
is not connected, that is when the commutative spectral triple is a direct
sum. \\
 \indent vi) In our definition of reducibility, the Dirac operator is not
supposed to leave the subspace $\hh_0$ invariant. Our definition is adapted
 to the use we will make of spectral triples: the fluctuations  under algebra
automorphisms properly lifted to the Hilbert space promote the Dirac
operator to a dynamical variable and we are interested in its dynamics, the
spectral action \cite{cc}.

\end{rem} 

Since we are mainly interested in finite or 0-dimensional triples, we only
recall the definition for this case  and also restricting ourselves to the real
and
$S^0$-real triples
\cite{real,grav}:

\begin{defn} A real, $S^0$-real, finite spectral triple is given by
($\Aa,\hh,\dd, $ $J,\epsilon,\chi$)  with a finite dimensional real algebra
$\Aa $, a faithful representation
$\rho$ of $\Aa$ on  a finite dimensional complex Hilbert space $\hh$. Four
additional operators are defined on
$\hh$: the Dirac operator
$\dd$ is selfadjoint, the real structure $J$ is antiunitary, and the $S^0$-real
structure $\epsilon$ and the chirality
$\chi$ are both unitary involutions. These operators  satisfy:
\bb
\hspace{-0.1cm}\bullet \hspace{2cm} J^2=1,\qq [J,\dd]=[J,\chi]=[\epsilon,\chi]=[\epsilon,\dd]=0, \qq \epsilon
J=-J \epsilon,\qq\dd\chi =-\chi \dd, \cr
[\chi,\rho(a)]=[\epsilon,\rho(a)]=[\rho(a),J\rho(b)J^{-1}]=
[[\dd,\rho(a)],J\rho(b)J^{-1}]=0, \forall a,b \in \Aa.
\eee
$\bullet$ The chirality
 can be written as a finite sum $\chi =\sum_i\rho(a_i)J\rho(b_i)J^{-1}.$ 
This condition is called orientability.\\ 
$\bullet$ The intersection form
$\cap_{ij}:=\T(\chi \,\rho (p_i) J \rho (p_j) J^{-1})$ is non-degenerate,
$\rm{det}\,\cap\not=0$. The
$p_i$ are minimal rank projections in $\Aa$. This condition is called
{\it Poincar\'e duality}.
\end{defn}

With the help of the projectors $(1\pm \chi )/2$ and
$(1\pm\epsilon )/2$,  the Hilbert space is decomposed as  
\bb
\hh=\hh_L\op\hh_R\op\hh_L^c\op\hh_R^c \, . \label{espacedehilbert}
\ee
The two first
components  correspond in physics to particles, $\epsilon =1$, the last two 
correspond to antiparticles, $\epsilon =-1$. We use the convention where
left-handed spinors have negative  and right-handed spinors have
positive chirality.

If we denote by $\rho _L$ the restriction of $\rho $ to
$\hh_L$, ... and by $\rho _R^c$ the restriction to $\hh_R^c$, we will
always write the representation in the following form
\bb \label{def de la rep}
\rho:=
\begin{pmatrix}
\rho_L & 0 & 0 & 0 \\ 0 & \rho_R & 0 & 0 \\  0 & 0 & \overline{ \rho_L^c}  &
0 \\ 0 & 0 & 0 & \overline{ \rho_R^c}
\end{pmatrix}.
\ee
With respect to the decomposition (\ref{espacedehilbert}) of $\hh$, 
$\dd$ has the form
\bb \dd=\pp{0&\mm&0&0\\
\mm^*&0&0&0\\ 0&0&0&\overline{\mm}\\
0&0&\overline{\mm^*}&0}. \label{opdirac}\ee
Note that the $S^0$-reality of internal spaces is equivalent to the absence of
Majorana-Weyl spinors. These are possible in 0-dimensional spaces, but
impossible in 4- and (1+3)-dimensional spaces $M$.

\section{Krajewski diagrams}

Krajewski and Paschke \& Sitarz have classified all finite, thus
0-dimensional, real spectral triples \cite{class,Kraj}. Let us summarize this
classification for the $S^0$-real case using Krajewski's diagrammatic
language.

\subsection{Conventions and multiplicity matrices}

\hspace{0.45cm} $\bullet$ The algebra: it is a finite sum of $N$ simple algebras,
$\Aa =
\oplus_{i=1}^{N} M_{n_i}(\kk_i)$ and $\kk_i=\rr,\cc,\hhh$ where $\hhh$
denotes the quaternions.

$\bullet$ The representation: let us start with the easy case, $\kk=\rr,
\hhh$ in all components of the algebra. The algebras
$M_n(\rr)$ and
$M_n(\hhh)$ only have  one irreducible representation, the fundamental
one on $\cc^{(n)}$, where $(n)=n$ for $\kk=\rr$ and
$(n)=2n$ for $\kk=\hhh$. Therefore
$\rho
$ is of the form
\bb \rho(\oplus_{i=1}^N a_i):=(\oplus_{i,j=1}^N a_i \otimes 1_{m_{ji}} \otimes 1_{(n_j)})\ 
\op\ ( \oplus_{i,j=1}^N 1_{(n_i)} \otimes 1_{m_{ji}} \otimes
\overline{a_j} ).\label{kranot}\ee
The multiplicities $m_{ij}$ are non-negative integers and we denote by
$1_n$ the $n\times n$ identity matrix and set by convention $1_0:=0$.
 At the same time the real structure $J$ permutes the two main
summands and complex conjugates them, while the $S^0$-real structure 
and the chirality read
\bb \epsilon &=&(\oplus_{i,j=1}^N 1_{(n_i)} \otimes 1_{m_{ij}} \otimes
1_{(n_j)})\
\op\ (\oplus_{i,j=1}^N 1_{(n_i)} \otimes (-1)1_{m_{ji}} \otimes 1_{(n_j)}),\\
\chi&=&(\oplus_{i,j=1}^N 1_{(n_i)} \otimes \chi_{ji}1_{m_{ji}} \otimes
1_{(n_j)})\
\op\ (\oplus_{i,j=1}^N 1_{(n_i)} \otimes \chi_{ji}1_{m_{ji}} \otimes
1_{(n_j)}),\ee
where $\chi_{ij}=\pm 1$ according to our previous convention 
on left-(right-)handed spinors.

We define the {\it multiplicity matrix} $\mu \in M_N(\zz)$ such
that $\mu _{ij}:=\chi _{ij} \, m_{ij}$. There are $N$ minimal projectors in
$\Aa$, each of the form
$p_i=0 \op  \cdots \oplus 0\op \rm{diag}(1_{(1)},0,...,0) \op 0\oplus \cdots
\op 0$. With respect to the basis $p_i/(1)$, the matrix of the
intersection form is $\mu +\mu ^T$. 

If the algebra has summands with $\kk=\cc$, things get more complicated.
Indeed $M_n(\cc)$ has two non-equivalent irreducible representations,
the fundamental one and its complex conjugate, so we change
 (\ref{kranot}) into
\bb
\rho(\oplus_{i=1}^N a_i):=(\oplus_{i,j=1;\alpha_i,\alpha_j}^N a_{i\alpha_{i}} \otimes
1_{m_{j\alpha_{j}i\alpha_{i}}} \otimes 1_{(n_j)})\ 
\op\ ( \oplus_{i,j=1}^N 1_{(n_i)} \otimes 1_{m_{j\alpha_ji\alpha_i}} \otimes
\overline{a_{j\alpha_{j}}} ).
\label{kranotcomplex}
\ee
where $\alpha_i=1$ when $a_i \in M_{n_i}(\kk), \kk=\rr,\hhh$ and $\alpha_i=1,2$ when $a_i \in
M_{n_i}(\cc)$, and $a_{i1}:=a_i,a_{i2}:=\overline{a_i}$.

\noindent Therefore the
multiplicity matrix is an integer valued square matrix of size equal to the number of
summands with $\kk=\rr$ and $\hhh$ plus two times the number of
summands with $\kk=\cc$ and decomposes into $N^2$ submatrices of size
$1\times 1$, $2\times 2$, $1\times 2$ and $2\times 1$.
 For example $\Aa=M_n(\cc)\op M_m(\cc)\op M_q(\rr)\owns (a,b,c)$ has a
$5\times 5$ multiplicity matrix. Let us label its rows and columns with
algebra elements:
\bb
\mu = \begin{array}{ccc}{\left(\begin{array}{ccc}\mu _{aa}&\mu
_{ab}&\mu _{ac}\\\mu _{ba}&\mu _{bb}&\mu _{bc}\\
\mu _{ca}&\mu
_{cb}&\mu_{cc}
\end{array}\right)}&
\begin{array}{c}{}^{a}_{\bar a}\\{}^{b}_{\bar b}\\c \end{array}\\
\hspace{-0.3cm}  a  \, \bar a \hspace{0.5cm} b \, \bar b   \hspace{0.5cm} c
&& \end{array}  \hspace{-0.5cm}.
\eee
If both entries $\mu _{ij}$ and $\mu _{ji}$ of the multiplicity
matrix are non-zero, then they must have the same sign.

The nonvanishing entries within each submatrix $1\times 2$ or
$2\times 1$, like $\mu _{ca}$ or
$\mu _{ac}$, must have the same sign, while the signs of the
nonvanishing entries in each
$2\times 2$ submatrix, e.g. $\mu _{aa}$ or $\mu _{ab}$ must be checker
board like: $\pp{+ & - \\ - & + }$ or $\pp{ - & + \\ + & - }$. 

The {\it contracted multiplicity matrix} $\hat\mu $ 
is the $N\times N$ matrix constructed from
$\mu $ by replacing each of the previous submatrices in $\mu
$ by the sum of the entries of the submatrix. 

$\bullet$ Poincar\'e duality: The last condition to be satisfied by the
multiplicity matrix reflects the Poincar\'e duality. With respect to the basis
$p_i/(1)$ introduced above, $(1)=1$ for $\kk=\rr$ and $\cc$, $(1)=2$ for
$\kk=\hhh$, the matrix of the intersection form is $\hat\mu +\hat\mu ^T$.
Therefore we must have $\det(\hat\mu +\hat\mu ^T)\not=0$.

$\bullet$ The Dirac operator: The components of the (internal) Dirac
operator are represented by horizontal or vertical lines connecting two
nonvanishing entries of opposite signs in the multiplicity matrix $\mu $
and we will orient them from plus to minus. Each arrow represents a
nonvanishing, complex submatrix in the Dirac operator: For instance 
$\mu_{ij}$ can be linked to $\mu_{ik}$ or $\mu_{kj}$ by  
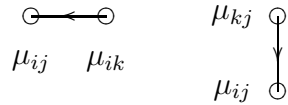
\begin{center}\begin{tabular}{cc}
\rxy{
,(0,0)*\cir(0.7,0){}
,(5,0)*\cir(0.7,0){}
,(5,0);(0,0)**\dir{-}?(.6)*\dir{>}
,(0,-3)*{\mu_{ij}}
,(5,-3)*{\mu_{ik}}
}
&\;\;\;\;\;
\rxy{
,(0,0)*\cir(0.7,0){}
,(0,-5)*\cir(0.7,0){}
,(0,0);(0,-5)**\dir{-}?(.6)*\dir{>}
,(-3,0)*{\mu_{kj}}
,(-3,-5)*{\mu_{ij}}
}
\end{tabular} \end{center}
and these arrows represent respectively submatrices of $\mm$ in $\dd$ of 
type $M\otimes 1_{(n_i)}$ with $M$ a complex $(n_j)\times(n_k)$ matrix
and $1_{(n_j)}\otimes M$  with $M$ a complex $(n_i)\times(n_k)$ matrix.

\noindent The requirement of 
non-degeneracy of a spectral triple means that every nonvanishing
entry in the multiplicity matrix
$\mu $ is touched by at least one arrow.

$\bullet$ Convention for the diagrams: We will see that (for sums of up to
three simple algebras) irreducibility implies that most entries of $\mu $
have an absolute value less than or equal to two. So we will use a {\it simple
arrow} to connect plus one to minus one and {\it double arrows} to
connect plus one to minus two or plus two to minus one (Figure 1.)
\begin{center}
\begin{tabular}{ccc}
\rxy{
,(0,0)*\cir(0.7,0){}
,(5,0)*\cir(0.7,0){}
,(5,0);(0,0)**\dir{-}?(.6)*\dir{>}
,(0,-3)*{-1}
,(5,-3)*{+1}
}
&
\;\;\;\;\;
\rxy{
,(20,0)*\cir(0.7,0){}
,(20,0)*\cir(0.4,0){}*\frm{*}
,(15,0)*\cir(0.7,0){}
,(20,0);(15,0)**\dir2{-}?(.6)*\dir2{>}
,(20,-3)*{+1}
,(15,-3)*{-2}
}
&
\;\;\;\;\;
\rxy{
,(35,0)*\cir(0.7,0){}
,(30,0)*\cir(0.7,0){}
,(30,0)*\cir(0.4,0){}*\frm{*}
,(35,0);(30,0)**\dir2{-}?(.6)*\dir2{>}
,(35,-3)*{+2}
,(30,-3)*{-1}
}
\\
\\
& Fig. 1&
\end{tabular}
\end{center}
Our arrows always point from plus, that is right chirality, to minus, that is
left chirality.
For a given algebra, every spectral triple is encoded in its
multiplicity matrix which itself is encoded in its Krajewski diagram, a field
of arrows. In our conventions, for particles, $\epsilon =1$, the column label
of the multiplicity matrix indicates the representation, the row label
indicates the multiplicity. For antiparticles, the row label
of the multiplicity matrix indicates the representation, the column label
indicates the multiplicity.

\noindent Every arrow comes with three algebras:  
Two algebras that localize its end
points, let us call them {\it right and left algebras} 
and a third algebra that localizes the arrow, let us call it {\it colour
algebra}.  For example for the arrow
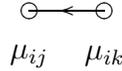
\bb\rxy{
,(0,0)*\cir(0.7,0){}
,(5,0)*\cir(0.7,0){}
,(5,0);(0,0)**\dir{-}?(.6)*\dir{>}
,(0,-3)*{\mu _{ij}}
,(5,-3)*{\mu _{ik}}}\eee
the left algebra is $\Aa _j$, the right algebra is $\Aa_k$ and the colour
algebra is $\Aa_i$.

\noindent The {\it circles} in the diagrams only intend to guide the
eye. A {\it black disk} on a double arrow indicates that the coefficient of the
multiplicity matrix is plus or minus one at this location, ``the two arrows
are joined at this location''.  For example the the following arrows
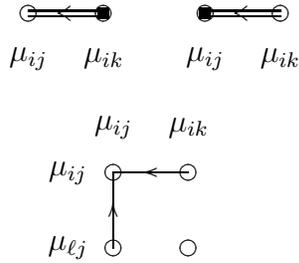
\begin{center}
\begin{tabular}{cc}
\rxy{
,(20,0)*\cir(0.7,0){}
,(20,0)*\cir(0.4,0){}*\frm{*}
,(15,0)*\cir(0.7,0){}
,(20,0);(15,0)**\dir2{-}?(.6)*\dir2{>}
,(20,-3)*{\mu_{ik}}
,(15,-3)*{\mu_{ij}}
}
&\qq
\rxy{
,(35,0)*\cir(0.7,0){}
,(30,0)*\cir(0.7,0){}
,(30,0)*\cir(0.4,0){}*\frm{*}
,(35,0);(30,0)**\dir2{-}?(.6)*\dir2{>}
,(35,-3)*{\mu_{ik}}
,(30,-3)*{\mu_{ij}}
}
\end{tabular}
\end{center}
\bb
\rxycc{0.7}{
,(10,-5);(5,-5)**\dir{-}?(.6)*\dir{>}
,(5,-10);(5,-5)**\dir{-}?(.6)*\dir{>}
}
\eee
represent respectively submatrices of 
$\mm$ of type $$\pp{M_1&M_2} \otimes 1_{(n_i)}\qq
{\rm and} \qq \pp{M_1\cr M_2} \otimes 1_{(n_i)}$$  
with $M_1,M_2$ of size $(n_j)\times(n_k)$ or 
in the third case, a matrix of type $\pp{M_1\otimes 1_{(n_i)}&1_{(n_j)}
\otimes M_2}$  where $M_1$ and $M_2$ are of size $(n_j)\times(n_k)$ and
$(n_i)\times(n_\ell)$.

According to these rules, we can omit the number 
$\pm1,\pm2$ under the arrows like in Figure 2, since they are now redundant. 
\vspace{1\baselineskip}

Let us give a few examples of the explicit form of the spectral triple
associated to a given Krajewski's diagram: \\  Take the algebra
$\Aa=\hhh\op M_3(\cc)\owns (a,b)$ with the first diagram of Figure 2. 
\begin{center}
\begin{tabular}{ccc}
\rxyzz{0.7}{
,(5,-5);(10,-5)**\dir{-}?(0.4)*\dir{<}
}
&
\;\;\;\;\;
\rxyzz{0.7}{
,(5,-5)*\cir(0.4,0){}*\frm{*}
,(5,-5);(10,-5)**\dir2{-}?(0.4)*\dir2{<}
}
&
\;\;\;\;\;
\rxyzz{0.7}{
,(5,-5);(10,-5)**\dir{-}?(0.4)*\dir{<}
,(5,-5);(5,-10)**\dir{-}?(0.4)*\dir{<}
}
\\
\\
&Fig. 2&
\end{tabular}
\end{center}
Then the multiplicity matrix and its contraction are 
\bb
\mu=\pp{-1&1&0 \cr 0&0&0 \cr 0&0&0},\qq
\hat\mu=\pp{-1&1 \cr 0&0 }.
\eee
Using (\ref{def de la rep}), its representation is, up to
unitary equivalence  
\bb
\rho_L (a,b)=a \ot 1_2,\, \rho _R (a,b) =b \ot 1_2,\, \rho_L^c (a,b)=1_2\ot a,\,
\rho _R^c(a,b)=1_3\ot a.
\eee The Hilbert space is
\bb
\hh=\cc^4 \oplus \cc^6 \oplus \cc ^4 \oplus \cc^6.
\eee
In its Dirac operator (\ref{opdirac}), $\mm=M\ot 1_2$, 
where $M$ is a nonvanishing complex $2\times 3 $
matrix.

\noindent Real structure, $S^0$-real structure and chirality are given by (cc stands for
complex conjugation)
\bb J=
\begin{pmatrix} 0&1_{10}\\ 1_{10}&0
\end{pmatrix}
\circ {\rm cc},\qq
\epsilon =
\begin{pmatrix} 1_{10}&0\\ 0&-1_{10}
\end{pmatrix} ,\qq \chi =
\begin{pmatrix}-1_4&0&0&0\\ 0&1_6&0&0\\ 0&0&-1_4&0\\ 0&0&0&1_6
\end{pmatrix} .\eee The first tensor factor in $a\ot 1_2$ concerns
particles, the second concerns antiparticles denoted by
$\cdot^c$.  The antiparticle representation is read from the transposed
multiplicity matrix.

The second diagram of Figure 2 yields 
\bb
\mu=\pp{-1&2&0 \cr 0&0&0 \cr 0&0&0},\qq 
\hat\mu=\pp{-1&2 \cr 0&0},
\eee
and its spectral triple reads:
\bb
\rho _L(a,b)=a\ot 1_2,\qq \rho _R(a,b)=\pp{b&0\\ 0& b}\ot 1_2,
\nonumber\\[2mm]
\rho _L^c(a,b)=1_2\ot a,\qq
\rho _R^c(a,b)=\pp{1_3&0\\ 0&1_3}\ot a, \cr \cr
\mm=\pp{M_1&M_2}\ot 1_2,\, M_1\ {\rm and}\ M_2
\ {\rm of \ size}\ 2\times 3,
\eee
\vspace{-0.9cm}
\bb
 J=\pp{0&1_{16}\\ 1_{16}&0}\circ {\rm cc},\qq
\epsilon =\pp{1_{16}&0\\ 0&-1_{16}},\qq \chi =\pp{-1_4&0&0&0\\
0&1_{12}&0&0\\ 0&0&-1_4&0\\ 0&0&0&1_{12}}.
\eee 
In Krajewski's notations, equation (\ref{kranot}), we would have
written
$\rho _R(a,b)=b\ot 1_2\ot 1_2$, the middle factor showing the entry of the
multiplicity matrix $\mu $.

Finally, still for the same algebra, let us consider the third diagram of
Figure 2. It gives
\bb
\mu=\pp{-1&1&0 \cr 1&0&0 \cr 0&0&0},\qq
\hat\mu=\pp{-1&1 \cr 1&0 },
\eee
and 
\bb
\rho _L(a,b)=a\ot 1_2,\qq \rho _R(a,b)=\pp{b\ot 1_2&0\\ 0& a\ot 1_3},
\nonumber \\[2mm]
\rho _L^c(a,b)=1_2\ot a,\qq
\rho _R^c(a,b)=\pp{1_3\ot a&0\\ 0&1_2\ot b},\cr \cr
\mm=\pp{M_1\ot 1_2&1_2\ot M_2},\qq M_1\ {\rm and}\ M_2
\ {\rm of \ size}\ 2\times 3.
\eee  In the three above examples all arrows have left algebra $\hhh$,
right algebra $M_3(\cc)$ and colour algebra
$\hhh$.
  The numerous examples below should allow the reader to get familiar
with the translation between diagrams and triples.

\subsection{Multiplicity matrix and irreducibility}

Our work is based on the following lemma indicating that a classification 
of irreducible spectral triples is possible.
\begin{lem}  i) The direct sum of multiplicity matrices is again a
multiplicity matrix describing the direct sum of spectral triples. \\ ii) For a
given algebra there is only a finite list of multiplicity matrices describing
irreducible triples.
\end{lem}
\begin{proof} i) is obvious.

ii) Given an algebra $\Aa$, denote by $\sss$ the set of multiplicity 
matrices $\mu \in M_n(\zz)$ ($n$ is determined by $\Aa$) 
associated to irreducible spectral triples. Define a partial order in $M_n(\zz)$ by 
$\mu \geq \nu$ when $\mu_{ij}$ and $\nu_{ij}$ have the same sign and 
$|\mu _{ij}|\geq|\nu_{ij}|$ for all $i,j=1,\ldots, n$. 
The interest of this order is that for two different
multiplicity matrices 
$\mu,\nu$ such that $\mu \geq \nu$ and $\nu \in \sss$, $\mu$ corresponds 
to a reducible triple.

To prove that card($\sss)< \infty$, we first identify $M_n(\zz)$ with $\zz^{n^2}$. We
denote by 
$\{e_i\}_{i \in \{1, \ldots, n^2\}}$ be the canonical basis of $\zz^{n^2}$ 
and $\sss^+:=\sss \cap \nn^{n^2}$.
We now prove that card($\sss^+ )< \infty$, which is sufficient.

Assume card($\sss^+ ) = \infty$. Then there exists at least one direction 
(say along $e_1$) such that sup$\{(\mu )_1 \vert \mu \in \sss^+ \}=\infty$. 
Suppose now that in each hyperplane defined by $me_1, m\in\nn$, all
points 
$\mu \in \sss^+$ remain uniformly bounded, 
$\mu_i<B,i =2,\ldots, n^2$. This means that there exists  an infinite family of
points in an hypertube parallel to $e_1$. Necessarily,  there exists an
infinite subfamily $\{\mu_k\}_{k\in \nn} \subset \sss^+$ of points in the 
hypertube which are aligned: $(\mu_k)_i=(\mu_{k+1})_i, \forall i\neq1$
and 
$(\mu_k)_1 < (\mu_{k+1})_1$. But this is impossible since 
$\mu_k \leq \mu_{k+1}$ cannot happen. As a consequence, there exists  a
second direction (say along $e_2$ after renumbering) where the
intersection  of successive hyperplanes along $e_1$ and $e_2$ contain an
infinite family of points of $\sss^+$. By induction, the same reasoning in
each direction 
$e_i$ implies that there exits an infinite family $\{\mu_k\}_k$ of points in
$\sss^+$  with increasing vectors: $(\mu_k)_i < (\mu_{k+1})_i, \forall i$.
Again  this yields the contradiction $\mu_k \leq \mu_{k+1}$ and 
$\mu_{k+1} \notin \sss^+$.
\end{proof}

\section{Fluctuations and dynamical non-degeneracy}

The aim of this work is two-fold. First, we work out all irreducible real,
$S^0$-real diagrams for algebras with one, two and three simple
summands. Second, we give all associated
spectral triples that are `dynamically non-degenerate'. By this we mean
the following. The spectrum of the (internal) Dirac operator $\dd$ is
always degenerate: all nonvanishing eigenvalues come in pairs of
opposite sign due to the chirality that anticommutes with $\dd$. All
eigenvalues appear twice due to the real structure that commutes with
$\dd$. There is a further degeneracy, two-fold in the example above,
$\mm=M\ot 1_2$, that comes from the first order axiom. Let us call it
{\it colour degeneracy}. It is absent if and only if the colour algebras of
all arrows are commutative. Of course, these three degeneracies survive
the fluctuations of the Dirac operator and the minimization of the
Higgs potential. By dynamical non-degenerate we mean (see precise
definition below) that no minimum of the Higgs potential has
degeneracies other than the above three. The first two degeneracies
survive {\it quantum fluctuations} as well. We also want the colour
degeneracies to be protected from quantum fluctuations. A natural
protection is unbroken gauge invariance, a requirement that we will
include into the definition of dynamical non-degeneracy.

Except for complex conjugation in $M_n(\cc)$ and permutations of
identical summands in the algebra $\Aa=\Aa_1\op\Aa_2\op ...\op\Aa_N$,
every algebra automorphism
$\sigma
$  is inner, $\sigma (a)=uau^{-1}$ for a unitary $ u\in U(\Aa)$. Therefore
the connected component of the automorphism group is
Aut$(\Aa)^e=U(\Aa)/(U(\Aa)\cap{\rm Center}(\Aa))$. Its lift to the Hilbert
space \cite{real}
\bb L(\sigma )=\rho (u)J\rho (u)J^{-1}\eee is multi-valued. 

The {\it fluctuation $\ddf$} of the Dirac operator $\dd$ is given by a
finite collection $f$ of real numbers 
$r_j$ and algebra automorphisms $\sigma _j\in{\rm Aut}(\Aa)^e$ such
that
\bb
\ddf :=\sum_j r_j\,L(\sigma _j) \, \dd \, L(\sigma_j)^{-1},\qq r_j\in\rr,\
\sigma _j\in{\rm Aut}(\Aa)^e.
\eee
The fluctuated Dirac operator $\ddf$ is often denoted by $\varphi $, the
`Higgs scalar', in the physics literature.  We consider only fluctuations
with real coefficients since 
$\ddf$ must remain selfadjoint.

To avoid the multi-valuedness in the fluctuations, we allow the entire
unitary group viewed as a (maximal) central extension of the
automorphism group. We will come back to minimal central extensions in
another work.

An almost commutative geometry is the tensor product of a finite
noncommutative triple with an infinite, commutative spectral triple. By
Connes' reconstruction theorem \cite{grav} we know that the latter comes
from a Riemannian spin manifold, which we will take to be any
4-dimensional, compact, flat manifold like the flat 4-torus.  The spectral
action of this almost commutative spectral triple reduced to the finite part
is a functional on the vector space of all fluctuated, finite Dirac operators:
\bb V(\ddf )= \lambda\  \T\!\left[ (\ddf )^4\right] -\textstyle{\frac{\mu ^2}{2}}\
\T\!\left[
(\ddf) ^2\right] ,\eee where $\lambda $ and $\mu $ are positive constants
\cite{cc,kraj2}.
The spectral action is invariant under lifted automorphisms and even
under the unitary group $U(\Aa)\owns u$,
\bb V( [\rho (u)J\rho (u)J^{-1}] \, \ddf \, [\rho (u)J\rho
(u)J^{-1}]^{-1})=V(\ddf),\eee and it is bounded from below.
 Our task is to find the minima $  \ddfm $ of this action,
their spectra and their  {\it little groups}
\bb G_\ell:=\left\{ u \in U(\Aa)),\ [\rho (u)J\rho
(u)J^{-1}] \,
\ddfm \, [\rho (u)J\rho (u)J^{-1}]^{-1}=\ddfm \right\} .\eee

\begin{defn} The irreducible
 spectral triple $(\Aa,\hh,\dd)$ is {\it dynamically non-degenerate} if all
minima $\ddfm$ of the action $V(\ddf)$ define a non-degenerate spectral
triple $(\Aa,\hh,\ddfm )$ and if the spectra of all minima  have no
degeneracies other than the three kinematical degeneracies: left-right,
particle-antiparticle and colour. Of course in the massless case there is no
left-right degeneracy. We also suppose that the colour degeneracies are
protected by the little group. By this we mean that all eigenvectors of
$\ddfm$ corresponding to the same eigenvalue are in a common orbit of
the little group (and scalar multiplication and charge conjugation).
\end{defn}
In physicists' language this last requirement means noncommutative
colour groups are unbroken. It ensures that the corresponding mass
degeneracies are protected from quantum corrections.

\section{Statement of the result}

The main result of this work is the following

\begin{thm}  The sum of simple algebras,
$\Aa=\bigoplus_{i=1}^N \Aa_i$ with $N=1,2,3$ admits a finite, real,
$S^0$-real,
 irreducible and dynamically non-degenerate spectral triple if and only if
it is in this list, up to a reordering of the  summands:
\begin{center}
\begin{tabular}{|c|c|c|}
\hline &&\\
 $N=1$&$N=2$&$N=3$\\[1ex]
\hline &&\\ &${\bf 1}\op{\bf 1}$&${\bf 1}\op{\bf 1}\op\ccc$\\[1ex]
  {\rm void}&&${\bf 1}\op{\bf 1}\op {\bf 1}$\\[1ex]
&${\bf 2}\op{\bf 1}$&${\bf 2}\op{\bf 1}\op\ccc$\\[1ex]
 &&${\bf 2}\op{\bf 1}\op {\bf 1}$\\[1ex]
\hline
\end{tabular}
\end{center} Here $\bf 1$ is a short hand for $\rr$ or $\cc$ and $\bf 2$
for $M_2(\rr)$,
$M_2(\cc)$ or
$\hhh$.\\ The `colour' algebra
$\ccc$ is any simple algebra and has two important constraints:\\ i) Its
representations on corresponding left- and right-handed subspaces of
$\hh$ are identical (up to possibly different multiplicities). \\ ii) The
Dirac operator $\dd $  is invariant under $U(\ccc)$,
\bb \rho (1,1,w) \, \dd \, \rho (1,1,w)^{-1}=\dd,\qq {\rm for \ all}\ w\in
U(\ccc).\eee This implies that the unitaries of ${}\ \ccc$ do not participate
in the fluctuations and are therefore unbroken, i.e. elements of the little
group.
\end{thm}

Let us emphasize that although the 4-dimensional `spacetime' manifold
$M$ used to define the almost commutative geometry does not
show up in this result, it is an important ingredient of the spectral action
and its asymptotic behavior. In particular the dimension of $M$ is linked
to the order of the polynomial $V$. Therefore our classification indeed concerns
4-dimensional, almost commutative geometries.

 We give in section \ref{exemple} an example of a reducible triple
which is dynamically non-degenerate and whose algebra is not in the
above list.

\section{One simple algebra}

From the classification \cite{class,Kraj}, we know that $\Aa=M_n(\cc)$ are
the only simple algebras to admit real spectral triples. Up to permutation of
$a$ and $\bar a$ (complex conjugation in $\Aa$),  up to permutation of
particles and antiparticles (reflection of the diagram with respect to the
main diagonal) and up to permutation of left- and right-handed particles
(changing the direction of all arrows), all real 
$S^0$-real and irreducible triples have Krajewski diagrams indicated in
Figure 3:

\begin{center}
\begin{tabular}{cccc}
\rxyc{0.7}{
,(10,-5)*\cir(0.4,0){}*\frm{*}
,(10,-5);(5,-5)**\dir2{-}?(0.6)*\dir2{>}
}
&
\;\;\;\;\;
\rxyc{0.7}{
,(10,-5);(5,-5)**\dir{-}?(0.6)*\dir{>}
,(10,-5);(10,-10)**\dir{-}?(0.6)*\dir{>}
}
&
\;\;\;\;\;
\rxyc{0.7}{
,(5,-5)*\cir(0.4,0){}*\frm{*}
,(10,-5);(5,-5)**\dir2{-}?(0.6)*\dir2{>}
}
&
\;\;\;\;\;
\rxyc{0.7}{
,(10,-5);(5,-5)**\dir{-}?(.6)*\dir{>}
,(5,-10);(5,-5)**\dir{-}?(.6)*\dir{>}
}
\\
\\
&\hspace{4cm} Fig. 3&
\end{tabular}
\end{center}
 Indeed, a Krajewski diagram must contain at least one arrow which can be put
into the position $\mu = \pp{-1&+1\\0&0}$ by use of the three above permutations.
However, alone this arrow does not fulfil Poincar\'e duality, $\hat \mu
+\hat
\mu ^T=0.$ There are four ways to add a second arrow. But Poincar\'e
duality can only be satisfied if the two arrows are joined in one point.
 Adding a third arrow makes the diagram corresponding to 
a reducible spectral triple.

\noindent The first diagram yields:
\bb \label{rep1}
\rho _L (a)=\pp{a&0\\ 0&a}\ot 1_n,\,
\rho _R(a)=\bar a\ot 1_n,\,
\rho _L^c(a)=1_n\ot\pp{a&0\\ 0&a},\,
\rho _R^c(a)=1_n\ot a,
\ee
\bb
\mm:=M \otimes 1_n:=\pp{M_1\\ M_2}\ot 1_n,\qq M_1,M_2\in M_n(\cc).
\eee Real structure, $S^0$-real structure and chirality are:
\bb J=\pp{0&1_{3n^2}\\ 1_{3n^2}&0}\circ \,{\rm cc},\,\,
\epsilon = \pp{1_{3n^2}&0\\ 0&-1_{3n^2}},\,
\chi =\pp{-1_{2n^2}&0&0&0\\ 0&+1_{n^2}&0&0\\ 0&0&-1_{2n^2}&0\\
0&0&0&+1_{n^2}}.\eee Let us write the fluctuations of $\dd$ as
\bb
\ddf :=\pp{0&\mmf &0&0\\
\mmf ^*&0&0&0\\ 0&0&0&\overline{\mmf} \\
0&0&\overline{{\mmf}^*}&0}.
 \label{opdiracfluc}
\ee
Here
\bb
\mmf =\Mf\ot1_n=\pp{\Mf_1 \\ \Mf_2}\ot 1_n.
\eee
Then
\bb
\Mf_1 =\sum_j r_j\,u _jM_1 u_j^{T},\,
\Mf_2 =\sum_j r_j\,u _jM_2 u_j^{T},\, {\rm so} \ 
\Mf =\sum_j r_j\,\pp{u _j&0\\ 0&u_j}
\pp{M_1\\ M_2} u_j^{T}.\label{image}
\eee and
\bb  V(\ddf )= \lambda\  \T\!\left[ \ddf ^4\right] -\frac{\mu^2}{2}\
\T\!\left[
\ddf ^2\right] = 4n \, (\lambda \ \T [({\Mf}^{\;*} \,
\Mf)^2]-\frac{\mu^2}{2}\ \T [{\Mf}^{\;*} \, \Mf]).
\eee The matrix $\Mf$ is of size $2n\times n $. Therefore the fluctuation
$\ddf
$ has at least $n$ vanishing eigenvalues each still coming with its
$n$-fold colour degeneracy and all triples with $n\ge 2$ are dynamically
degenerate since $\Mf \, {\Mf}^{\;*} \in M_{2n}(\cc)$ has at  least $n$
zero eigenvalues.

For $n=1$, all minima of the action $V$ are of the form
$|\Mfm_1|^2+|\Mfm_2|^2={\textstyle\frac{\mu ^2}{4\lambda}}$. But the
corresponding fluctuated Dirac operator has a nontrivial invariant
subspace in its kernel and the triple is degenerate. To make the subspace
explicit apply a unitary change of basis to set $\Mf _2$ to zero. In every
minimum the little group is $\zz_2$.

At this point, an overkill is instructive. We will show that the case
$n\ge 2$ also features dynamical degeneracy in the non-zero eigenvalues.

In general, the set of all possible fluctuations $\ddf $, i.e. the image under
the fluctuations (\ref{image}), is difficult to describe. However, the action
$V$ only depends on the positive
$n\times n$ matrix 
$C:={\Mf}^{\;*} \, \Mf$ and is a sum of $n^2$ polynomials of fourth order
in the matrix elements of $C$:
\bb V(C)= 4n \, (\,\sum_{i=1}^n(\lambda C_{ii}^2-{\textstyle\frac{\mu
^2}{2}}  C_{ii})\,+\,\sum_{i\not= j}\lambda |C_{ij}|^2).\ee
 If $C={\textstyle\frac{\mu ^2}{4\lambda }} 1_n$ is in this image then it is
the unique minimum in terms of the variable $C$.

To compute the minima of the action, we now distinguish cases:

{\bf 1:} At least one diagonal element of one of the two matrices is
non-zero:

After a suitable renumbering of the basis of the Hilbert space $\hh$, we
may assume
$(M_1)_{11} \not=0$. With a first fluctuation,
\bb r_1={\textstyle\frac{1}{2}} , r_2={\textstyle\frac{1}{2}} ,\qq u_1=1_n,\
u_2=\pp{1&0\\ 0&-1_{n-1}},\eee we obtain for $\Mf$ a  block diagonal
matrix with $2\times 2$ blocks. By means of the fluctuation
\bb r_1={\textstyle\frac{1}{2}} , r_2={\textstyle\frac{1}{2}} ,\qq u_1=1_n,\
u_2=\pp{1&0\\ 0&i1_{n-1}},\eee we isolate the (1,1) elements of $M_1$ and
$M_2$ and with $r_1=r_2=...=r_n=1$, $ u_1=1_n,$
\bb
 u_2=\pp{0&1&0\\1&0&0\\ 0&0&1_{n-2}},\
 u_3=\pp{0&0&1&0\\0&1&0&0\\ 1&0&0&0\\ 0&0&0&1_{n-3}},\
 u_4=\pp{0&0&0&1&0\\0&1&0&0&0\\ 0&0&1&0&0\\ 1&0&0&0&0\\0&
0&0&0&1_{n-4}},...\eee we distribute them over the entire diagonal
 obtaining
\bb \Mf=\pp{(M_1)_{11}\,1_n\\ (M_2)_{11}\,1_n}.\eee Then $C$ is a
non-vanishing multiple of the identity and a suitable multiple of the above
$\Mf$ is a minimum. The spectrum of this minimum $\ddfm$ has an
additional dynamical degeneracy and up to the sign, $\ddfm$ has one
single eigenvalue. The little group is
$G_\ell=O(n)$.  Note also that $\Mfm_1$ and $\Mfm_2$ are proportional,
$\Mfm_1=\alpha  \, \Mfm_2$, $\alpha  \in\cc$, so there is a unitary
change of basis in the Hilbert space $\hh$ such that $\Mfm_2=0$  in the
new basis and the spectral triple $(M_n(\cc),\hh,\ddfm)$ is degenerate.

{\bf 2:} All diagonal elements of $M_1$ and $M_2$ vanish but $M_1$ and
$M_2$ are not both skewsymmetric:

After a suitable renumbering, we may assume $(M_1)_{12}=\beta -
\gamma  $,  $(M_1)_{21}=\beta +\gamma  $,
$(M_1)_{11}=(M_1)_{22}=0$, with $\beta  \not=0$. As in case {\bf 1} we can
isolate this block. Fluctuating with
\bb r_1=1,\qq u_1 =\pp{1/\sqrt 2&-1/\sqrt 2&0\\ 1/\sqrt 2&1/\sqrt 2&0\\
0&0&1_{n-2}}, \eee we obtain $(\Mf_1)_{11}=-\beta  $ and conclude as in
case {\bf 1}.

{\bf 3:} $M_1$ and $M_2$ are skewsymmetric and linearly dependent:

Thus M can be written as $M=M_1 \otimes \pp{1 \\ \alpha}$, $\alpha \in
\cc$  and by a unitary change of basis, we may assume $\alpha=0$ that is
$M_2=0$  without changing the representation (\ref{rep1}). Then the
spectral triple is degenerate. Let us nevertheless finish this case. By
another unitary change of basis (or fluctuation), 
cf. appendix (\ref{antisym}), we put
\bb M_1=\pp{\pp{0&-\lambda _1\\ \lambda _1&0}&&&\\ &\pp{0&-\lambda
_2\\ \lambda _2&0}&&\\ &&\ddots&\\ &&&0},\eee where the zero in the
lower right corner concerns the case $n$ odd. Let us suppose that
$\lambda _1$ is not zero. By isolating the upper left block and by
distributing it over the entire diagonal we obtain a
${\Mf_1}_{ii}=\lambda_1$ for all $i$. If
$n$ is even, we then have the minimum $\hat C={\textstyle\frac{\mu
^2}{4\lambda }} 1_n$ for $V(C)$. Its spectrum is again completely
degenerate with little group
$G_\ell=USp({\textstyle\frac{n}{2}}).$ If $n$ is odd the spectrum of
$\ddfm$ contains 4 vanishing eigenvalues, all others being of same
absolute value and
$G_\ell =USp(\frac{n-1}{2})\times U(1).$

{\bf 4:} $M_1$ and $M_2$ are skewsymmetric and linearly independent:

Then $\Mf_1$ and $\Mf_2$ vary independently over all skewsymmetric
matrices (cf. Lemma A.2). For
$n=3$ the minimization can be done by direct calculation. All minima are
gauge equivalent to
\bb {}^{\hat f}M=\,{\textstyle\frac{\mu}{\sqrt{6\lambda}}}\, \pp{0&1&0\\
-1&0&0\\ 0&0&0\\ 0&0&1\\ 0&0&0\\ -1&0&0},\qq \hat C:={\Mfm}^{\;*} \,
\Mfm=\,{\textstyle\frac{\mu^2 }{{6\lambda }}}\, \pp{2&0&0\\ 0&1&0\\
0&0&1}.\eee Although the spectrum has a two-fold dynamical degeneracy
the little group is only $G_\ell=U(1)\owns$ diag$(e^{i\theta}
,e^{-i\theta},e^{-i\theta})$. For general $n$, we were unable to compute a
minimum explicitly, but we will show that its spectrum is dynamically
degenerate. To alleviate notations let us rescale variables, $\ddf
\rightarrow \frac{\mu}{\sqrt{2\lambda}} \,\ddf $. Then
 the spectral action reads for $C_i:={\Mf_i}^* \,\Mf_i$
\bb V(\Mf_1,\Mf_2)=\,{\textstyle\frac{n\mu ^4}{\lambda }}\, \left( {\rm
tr}[C_1^2] -{\rm tr}[C_1] +{\rm tr}[C_2^2] -{\rm tr}[C_2] +2 \, {\rm tr}[C_1C_2]
\right).
\eee All minima of this fourth order polynomial have vanishing partial
derivatives with respect to ${\Mf_1}^*$ and ${\Mf_2}^*$. These equations
read
\bb &\Mfm_1=2\,\Mfm_1\,{\Mfm_1}^*\,\Mfm_1
+\Mfm_1\,{\Mfm_2}^*\,\Mfm_2
+\Mfm_2\,{\Mfm_2}^*\,\Mfm_1,&\label{f1}\\
 &\Mfm_2=2\,\Mfm_2\,{\Mfm_2}^*\,\Mfm_2 +\Mfm_2 \,{\Mfm_1}^*
\,\Mfm_1 +\Mfm_1 \,{\Mfm_1}^* \,\Mfm_2.&\label{f2}
\eee Let us put $X:=\hat{C_1}$, $Y:=\hat{C_2}$ and $Z:={\Mfm_1}^* \,
\Mfm_2$. We multiply equation (\ref{f1}) from the left by $\Mfm_1$ and
likewise for (\ref{f2}) to get
\bb&X=2X^2+XY+ZZ^*,&\label{X}\\ &Y=2Y^2+XY+Z^*Z.&\label{Y}\eee
Subtracting  (\ref{X}) from its Hermitian conjugate, we have that $X$ and
$Y$ commute and subtracting (\ref{Y}) from (\ref{X}) we get
\bb 2(X-Y)(X+Y-{\textstyle\frac{1}{2}} 1_n)=[Z^*,Z].\label{XY}\eee
Multiplying (\ref{f2}) from the left with ${\Mfm_1}^*$  and multiplying
the Hermitian conjugate of (\ref{f1}) from the right with
$\Mfm_2$ , we have
\bb &Z=2ZY+ZX+XZ,&\\ &Z=2XZ+YZ+ZY.&
\eee Taking the difference, we find that $Z$ commutes with $X+Y$.

We are to show that the spectrum of $\hat C={\Mfm}^* \Mfm =X+Y$ is
degenerate: Let us suppose that it is non-degenerate. Take an orthonormal
basis of eigenvectors of $\hat C$. Then it is also an eigenbasis of $Z$
implying that $[Z,Z^*]=0$ and by (\ref{XY}) the eigenvalues $x_j$ and
$y_j$ of $X$ and $Y$ corresponding to the j-th basis  vector satisfy at least
one of the equations $x_j=y_j$, 
$x_j+y_j={\textstyle\frac{1}{2}} $. But as $\Mfm_1$ and $\Mfm_2$ are 
skewsymmetric, each eigenvalue of $X$ and $Y$ is doubly degenerate with
the exception of one vanishing eigenvalue if
$n$ is odd. This contradicts the non-degeneracy of $X+Y$.

The triple of the second diagram of figure 3 differs from the first only
with respect to representation and Dirac operator
\bb
\rho _L (a)=\pp{a&0\\ 0&\bar a}\ot 1_n,\qq
\rho _R(a)=\bar a\ot 1_n,\qq
\rho _L^c(a)=1_n\ot\pp{a&0\\ 0&\bar a},\qq
\rho _R^c(a)=1_n\ot a,
\eee
\bb 
\mm=\pp{ M_1\ot 1_n\\1_n\ot M_2},\qq M_1,M_2\in M_n(\cc).
\eee Again, non-degeneracy of the zero eigenvalue requires $n=1$ and all
minima, $|\Mfm_1|^2+|\Mfm_2|^2=\frac{\mu ^2}{4\lambda}$,
 have little group $\zz_2$. But now all minima of the action
$V$ are not gauge equivalent. Indeed when $\Mfm_1=\mu (4\lambda
)^{-1/2}$, $\Mfm_2=0$ and  Ker($\dd)={\rm Span} \{
{\hh_L}_2,{\hh_L^c}_2 \})$. Thus the eigenvector in the image of 
${\textstyle\frac{1}{2}}(1-\epsilon)$ associated to the zero eigenvalue of
the fluctuated   Dirac operator, which is in ${\hh_L}_2$, transforms under
$\rho (u)J\rho (u)J^{-1}$  as multiplication by $u^{-2}$, while it
transforms as multiplication by $u^2$ when
$\Mfm_2=\mu (4\lambda )^{-1/2}$. According to our definition the triple is
dynamically degenerate because in both cases, the eigenvectors define a
one-dimensional complex subspace invariant under $\Aa$ and contained
in the kernel of the fluctuated Dirac operator $\ddfm$.

The last two diagrams of Figure 3 are treated as the first two and yield the
same conclusions.

\section{Two simple algebras}

Again our starting point is the list, Figure 4, of all irreducible Krajewski
diagrams up to the three mentioned types of permutations and up to
permutations of the two algebras and disregarding any direct sum of two
diagrams from Figure 3.

\begin{center}
\begin{tabular}{cc}
\rxyb{0.7}{
,(15,-5);(5,-5)**\crv{(10,-8)}?(.6)*\dir{>}
}
&
\;\;\;\;\;
\rxyb{0.7}{
,(15,-5);(10,-5)**\dir{-}?(.6)*\dir{>}
}
\\
\\
\rxyb{0.7}{
,(20,-5)*\cir(0.4,0){}*\frm{*}
,(20,-5);(15,-5)**\dir2{-}?(.6)*\dir2{>}
}
&
\;\;\;\;\;
\rxyb{0.7}{
,(20,-5);(15,-5)**\dir{-}?(.6)*\dir{>}
,(20,-5);(20,-10)**\dir{-}?(.6)*\dir{>}
}
\\
\\
& \hspace{-4cm} Fig. 4
\end{tabular}
\end{center}

 Let $k$ and $\ell$ be the size of the matrices of $\Aa_1=M_n(\kk)\owns a$
and $\Aa_2=M_m(\kk)\owns b$. E.g.
$k=n$ for
$\Aa_1=M_n(\rr)$ or
$M_n(\cc)$ and $k=2n$ for $\Aa_1=M_n(\hhh).$

The first diagram of figure 4 yields:
\bb \rho _L(a,b) =a\ot 1_k,\qq
\rho _R(a,b)=b\ot 1_k,\qq
\rho _L^c(a,b)=1_k\ot a,\qq
\rho _R^c(a,b)=1_\ell\ot a,\eee
\bb \mm=M\ot 1_k,\qq M\in M_{k\times \ell}(\cc).\eee
$M$ is non-zero and we may assume $M_{11}\not= 0$. Except for the
$k$-fold colour degeneracy, we accept at most one zero eigenvalue of
$M^*M$. Therefore we must have $k=\ell$ or $k=\ell \pm 1$.  We assume
$\ell \leq k$. Let in (\ref{opdiracfluc})
\bb
\mmf =\Mf\ot1_k, \qq
\Mf =\sum_j r_j\,u _jM v_j^{-1},\qq u_j\in U(\Aa_1),\qq v_j\in U(\Aa_2).
\eee
If $k\ge 2, \ell>2$ or $k> 2, \ell\ge 2$,  we may isolate the upper
$2\times 2$ block by fluctuations: With the first fluctuation,
\bb r_1={\textstyle\frac{1}{2}} ,\ r_2={\textstyle\frac{1}{2}} ,\qq
u_1=1_k,\ v_1=1_\ell,\ u_2=\pp{1_2&0\\ 0&-1_{k-2}},\ v_2=\pp{1_2&0\\
0&-1_{\ell-2}},\eee
 we obtain for $M$ a block diagonal type matrix.
 By means of the fluctuation
\bb r_1={\textstyle\frac{1}{2}} ,\ r_2={\textstyle\frac{1}{2}} ,\qq
u_1=1_k,\ v_1=1_\ell,\ u_2=1_k,\ v_2=\pp{1_2&0\\ 0&-1_{\ell-2}}.\eee we
isolate the upper block. If $k=\ell= 2$ this step is empty.

Now we may distinguish cases.

{\bf 1:} $\Aa_1=M_n(\rr)$ or $M_n(\cc)$, $\Aa_2=M_m(\rr)$ or
$M_m(\cc)$: like above, we isolate $M_{11}$ and as in case {\bf 1} for one
algebra we distribute $M_{11}$ over the entire diagonal, obtaining thus
${\Mf}^* \, \Mf$ proportional to the identity. The spectrum of the
fluctuated Dirac operator $\ddf$ minimizing the action $V$ has an
$\ell$-fold dynamical degeneracy.

{\bf 2:} $\Aa_1=M_n(\hhh)$ , $\Aa_2=M_m(\hhh)$, define the fluctuation
\bb r_1={\textstyle\frac{1}{2}}, r_2={\textstyle\frac{1}{4}},
r_3={\textstyle\frac{1}{4}},\qq u_1=1_k,\ v_1=1_\ell,\ u_2=\pp{\pp{0&1\\
-1&0}&0\\ 0&1_{k-2}},
\eee
\bb v_2=\pp{\pm\pp{0&1\\ -1&0}&0\\ 0&1_{\ell-2}},\, u_3=\pp{\pp{0&i\\
i&0}&0\\ 0&1_{k-2}},\, v_3=\pp{\pm\pp{0&i\\ i&0}&0\\ 0&1_{\ell-2}},
\eee the upper block is proportional to $1_2$. The plus signs in $\pm$ are
used if $M_{11}=M_{22}$. We distribute the block over the diagonal and get
an
$\ell$-fold dynamical degeneracy (recall that $\ell \leq k$).

  {\bf 3:}
$\Aa_1=M_n(\rr)$ or $M_n(\cc)$,
$\Aa_2=M_m(\hhh)$. If $M_{12}=0$ we fluctuate with
\bb  r_1=r_2={\textstyle\frac{1}{4}}, r_3={\textstyle\frac{1}{2}} ,\qq
u_1=1_k,\ v_1=1_\ell,\ u_2=\pp{\pp{1&0\\ 0&-1}&0\\ 0&1_{k-2}} ,\ 
v_2=1_\ell,
\eee
\bb  u_3=\pp{\pp{0&1\\ 1&0}&0\\ 0&1_{k-2}} ,\ v_3=\pp{-\pp{0&-1\\
1&0}&0\\ 0&1_{\ell-2}},
\eee and obtain
\bb
\Mf=\pp{\pp{M_{11}&0\\ 0&-M_{11}}&0\\ 0&0}.
\eee If $M_{12}\not=0$ we fluctuate with
\bb  r_1=r_2={\textstyle\frac{1}{4}} ,
\ r_3={\textstyle\frac{1}{2}} ,\qq u_1=1_k,\ v_1=1_\ell,\ u_2=\pp{1_2&0\\
0&-1_{k-2}} , \ v_2=1_\ell,
\eee
\bb u_3=1_k ,\ v_3=\pp{\pp{e^{-i\theta }&0\\ 0&e^{i\theta }}&0\\
0&-1_{\ell-2}},\eee with $\theta :={\textstyle\frac{1}{2}} ({\rm
Arg}(M_{11})-{\rm Arg}(M_{12}))$ and obtain
\bb \Mf =\pp{\pp{e^{i\theta }M_{11}&e^{-i\theta }M_{12}\\ 0&0}&0\\
0&0}. \eee In both cases, we distribute the
$2\times 2$ block over the diagonal and achieve ${\Mf}\,^* \, \Mf$
proportional to the identity.

In all three cases, $\ell$ must be one to avoid dynamical degeneracy.

The second diagram of Figure 4 is treated in the same fashion. The last two
diagrams of Figure 4 have no `letter changing arrow', an arrow
connecting an
$a$ to a $b$. They are treated as the triples with one simple algebra: {\it
Counting neutrinos}, that is requiring at most one zero eigenvalue (up to a
possible colour degeneracy) yields $\Aa =\cc\op\cc$ and degeneracy.

Finally using the permutations we get the list of all irreducible,
dynamically non-degenerate triples with two algebras:

- There are the commutative triples, that is the two-point spaces,
$\Aa=\cc\op\cc\owns (a,b)$:
\bb
\rho (a,b)=\pp{a&0&0&0\\ 0&b&0&0\\ 0&0&\bar a&0\\ 0&0&0&\bar a},
\,\, {\rm and}\,{\rm in } \,\, \dd, \, \mm\in\cc.\label{com1}
\ee
There is a second one with the same algebra:
\bb \rho (a,b)=\pp{\bar a&0&0&0\\ 0&b&0&0\\ 0&0& \bar a&0\\
0&0&0&\bar a}.\label{com2}\ee
And there are the real versions,
$\Aa=\cc\op\rr,\ \rr\op\cc$ and
$\rr\op\rr$.

- The noncommutative triples have $\Aa=M_2(\cc)\op\cc\owns (a,b)$ with
four irreducible triples:
\bb
\rho (a,b)=\pp{a\ot 1_2&0&0&0\\ 0&b1_2&0&0\\ 0&0& 1_2\ot\bar a&0\\
0&0&0&\bar a},&&\mm=\pp{0\\ m}\ot 1_2,\ m\in\cc, \label{kin1} \\[2mm]
\rho (a,b)=\pp{\bar a\ot 1_2&0&0&0\\ 0&b1_2&0&0\\ 0&0& 1_2\ot \bar
a&0\\ 0&0&0&\bar a},&&\mm=\pp{0\\ m}\ot 1_2, \label{kin2} \\[2mm]
\rho (a,b)=\pp{a&0&0&0\\ 0&b&0&0\\ 0&0& \bar b 1_2&0\\ 0&0&0&\bar
b},&&\mm=\pp{0\\ m}, \label{kin3} \\[2mm]
\rho (a,b)=\pp{a&0&0&0\\ 0&\bar b&0&0\\ 0&0&\bar  b 1_2&0 \\
0&0&0&\bar b},&&\mm=\pp{0\\ m}\label{kin4}.
\ee
In all four cases, all
minima  $\ddfm$ are gauge equivalent to the Dirac operator $\dd$ with the
absolute value of $m$ fixed in terms of
$\lambda $ and $\mu $ and the little groups are
\bb G_\ell= U(1)\times U(1)\owns\left( \pp{e^{i\alpha }&0\\ 0&e^{-i\beta
}},e^{i\beta }\right).\eee
 In the two triples (\ref{kin1}) and (\ref{kin2}), 
the unitaries of the colour algebra, $M_2(\cc)$, are spontaneously broken,
they do not leave any minimum invariant,
$U(2)\not\subset G_\ell$. According to our definition these two triples are
dynamically degenerate.

- Then there are diverse real versions: replace the complex
$2\times 2$ matrices $M_2(\cc)$ by quaternions  $\hhh$ or by real
matrices $M_2(\rr)$ and/or  replace $\cc$ by
$\rr$. For the real forms, of course, we have no complex conjugations in
the representations. We summarize the little groups (arrows mean group
homomorphisms):
\bb \cc\op\cc\qq\supset& U(1)\times U(1)&\longrightarrow\qq\qq
U(1),\cr
\cc\op\rr\qq\supset& U(1)\times
\zz_2&\longrightarrow\qq\qq U(1),\cr
\rr\op\rr\qq\supset& \zz_2\times
\zz_2&\longrightarrow\qq\qq \zz_2,\cr M_2(\cc)\op\cc\qq\supset&
U(2)\times U(1)&\longrightarrow\qq U(1)\times U(1),\cr
M_2(\cc)\op\rr\qq\supset& U(2)\times
\zz_2&\longrightarrow\qq U(1)\times \zz_2,\cr
\hhh\op\cc\qq\supset& SU(2)\times U(1)&\longrightarrow\qq \qq
U(1),\cr
\hhh\op\rr\qq\supset& SU(2)\times
\zz_2&\longrightarrow\qq\qq\ \zz_2,\cr
M_2(\rr)\op\cc\qq\supset&O(2)\times U(1)&\longrightarrow\qq\ 
\zz_2\times \zz_2,\cr M_2(\rr)\op\rr\qq\supset& O(2)\times
\zz_2&\longrightarrow\qq \ \zz_2\times \zz_2 .\label{little} \ee
 The triples  (\ref{kin1}) or its commutative version (\ref{com1}),
(\ref{kin2}) or its commutative version (\ref{com2}), (\ref{kin3}), and
(\ref{kin4}) are represented by the four diagrams of figure 5.
\begin{center}
\begin{tabular}{cc}
\rxyb{0.7}{
,(20,-5);(5,-5)**\crv{(12.5,-9)}?(.6)*\dir{>}
}
&
\;\;\;\;\;
\rxyb{0.7}{
,(20,-5);(10,-5)**\crv{(15,-8)}?(.6)*\dir{>}
}
\\
\\
\rxyb{0.7}{
,(15,-15);(5,-15)**\crv{(10,-18)}?(.6)*\dir{>}
}
&
\;\;\;\;\;
\rxyb{0.7}{
,(20,-15);(5,-15)**\crv{(12.5,-19)}?(.6)*\dir{>}
}
\\
\\
&\hspace{-4cm} Fig. 5
\end{tabular}
\end{center}

\section{Three simple algebras}
\subsection{Proof for $N=3$}

So far, we found that all irreducible, dynamically non-degenerate triples
were associated to diagrams with letter changing arrows only, i.e. arrows
connecting two different algebras. These arrows are stable under
contraction of the multiplicity matrix. Therefore we start by constructing
all irreducible, contracted diagrams, Figure 6. In other words we neglect
the complex conjugate representations.

This list becomes exhaustive upon permutations of the three algebras
$\Aa_1=M_n(\kk)\owns a$, $\Aa_2=M_m(\kk)\owns b$,
$\Aa_3=M_q(\kk)\owns c$, upon permuting left and right, i.e. changing
the directions of all two or three arrows simultaneously, and upon
permutations of particles and antiparticles independently in every
connected component of the diagram.

Let  $k$, $\ell$, $p$ be the sizes of the matrices $a,$ $b,$ $c$.
\vspace{1\baselineskip}

{\bf Diagram 1} yields:
\bb \rho _L (a,b,c)=\pp{a\ot 1_k&0\\ 0&b\ot 1_\ell},&&
\rho _R(a,b,c)=\pp{b\ot 1_k&0\\ 0&c\ot 1_\ell},\cr \cr \cr
\rho _L^c(a,b,c)=\pp{1_k\ot a&0\\ 0&1_\ell\ot b},&&
\rho _R^c(a,b,c)=\pp{1_\ell\ot a&0\\ 0&1_p\ot b},\eee and
\bb \mm=\pp{M_1\ot 1_k&0\\ 0&M_2\ot1_\ell},\qq M_1\in M_{k\times
\ell}(\cc),\ M_2\in M_{\ell\times p}(\cc).\eee The fluctuations,
\bb\Mf_1  &=&\sum_j r_j\,u _jM_1 v_j^{-1},\qq u_j\in U(\Aa_1),\qq v_j\in
U(\Aa_2),\cr
\Mf_2  &=&\sum_j r_j\,v _jM_2 w_j^{-1},\qq w_j\in U(\Aa_3) ,\eee produce
two {\it decoupled} fields $\Mf_1$ and $\Mf_2$ as can be seen by applying
the fluctuation:
$ r_1={\textstyle\frac{1}{2}},$ $u_1=1_k,$ $v_1=1_\ell,$ $w_1=1_p$,
$ r_2={\textstyle\frac{1}{2}},$ $u_2=1_k,$ $v_2=1_\ell,$ $w_2=-1_p$.

Since the arrows $M_1$ and $M_2$ are disconnected, the action is a sum of
an action in $\Mf_1$ and of an action in $\Mf_2$. Proceeding as in the
preceding section we find that a minimum
$\Mfm_1$ has min$\{k,\ell\}$ eigenvalues
$\mu{(4\lambda})^{-\frac{1}{2}}$  and $|k-\ell|$ eigenvalues zero and
$\Mfm_2$ has min$\{\ell,p\}$ eigenvalues
$\mu{(4\lambda})^{-\frac{1}{2}}$ and $|\ell-p|$ eigenvalues zero. All
triples associated to the first diagram are therefore dynamically
degenerate.

For the same reason, we can discard {\bf diagrams 2, 3, 4, 6, } because they
also have two disconnected horizontal arrows not vertically aligned.
\vspace{1\baselineskip}

{\bf Diagram 5} yields:
\bb \rho _L (a,b,c)=\pp{a\ot 1_k&0\\ 0&b\ot 1_p},&&
\rho _R(a,b,c)=b\ot 1_k,\cr \cr \cr
\rho _L^c(a,b,c)=\pp{1_k\ot a&0\\ 0&1_\ell\ot c},&&
\rho _R^c(a,b,c)=1_\ell\ot a,\eee \vspace{-0.5cm}
\bb \mm=\pp{M_1\ot 1_k\\ 1_\ell\ot M_2},\qq M_1\in M_{k\times
\ell}(\cc),\ M_2\in M_{p\times k}(\cc).\eee Again the fluctuations,
\bb\Mf_1  &=&\sum_j r_j\,u _jM_1 v_j^{-1},\qq u_j\in U(\Aa_1),\qq v_j\in
U(\Aa_2),\cr
\Mf_2  &=&\sum_j r_j\,w _jM_2 u_j^{-1},\qq w_j\in U(\Aa_3) ,\eee produce
two  decoupled fields $\Mf_1$ and $\Mf_2$ but now the arrows are
connected and consequently, the action does not decouple,
\bb V(C_1,C_2) =  4k \, [\lambda \, {\rm tr} (C_1^2) -  {\textstyle\frac{1}{2}}
\mu ^2  \, {\rm tr} (C_1)] +4 \ell [\lambda {\rm tr} (C_2^2) -
{\textstyle\frac{1}{2}} \mu ^2 {\rm tr} (C_2)] +8 \lambda \,{\rm tr}
(C_1)\,{\rm tr} (C_2),
\eee
where $C_i:={\Mf_i}^* \, \Mf_i$.
 Let $x_1,x_2,...,x_\ell$ be the eigenvalues of $C_1$ and $y_1,y_2,...,y_k$ be
the eigenvalues of $C_2$. The action only depends on these variables and
in its minimum all $x_i$ are equal or vanish and all $y_i$ are equal or
vanish. The spectrum of the minimal Dirac operator $\ddfm$ contains at
most three non-vanishing numbers: $\sqrt x, \sqrt y, \sqrt{x+y}$
implying that
$k$ and $\ell$ are less than or equal to two.
 The fermionic mass matrix $\mm$ is of size $(k^2+\ell p)\times (k\ell)$.
To get at most one zero eigenvalue we must require $|k^2+\ell p-k\ell|\le 1$
implying $k=p=1$. For $\ell=1$ the minimum is at
$|\Mfm_1|^2+|\Mfm_2|^2=\frac{\mu ^2}{4\lambda}$, for $\ell= 2$ the
minimum is at $\Mfm_1 =0,$ $|\Mfm_2|^2=\frac{\mu ^2}{4\lambda}$. In
both cases, $\Mfm_1 =0,$ the triple is degenerate in the sense that the
Dirac operator has an invariant subspace in the kernel.
\vspace{1\baselineskip}

{\bf Diagram 7} falls in the same way.
\vspace{1\baselineskip}

{\bf Diagram 8} yields the representations
\bb \rho _L(a,b,c) =\pp{a\ot 1_k&0&0\\ 0&c\ot 1_k&0\\ 0& 0&b\ot 1_p},&&
\rho _R(a,b,c)=\pp{b\ot 1_k&0\\ 0&c\ot 1_p},\cr \cr \cr
\rho _L^c(a,b,c)=\pp{1_k\ot a&0&0\\ 0&1_p\ot a&0\\ 0& 0&1_\ell\ot c},&&
\rho _R^c(a,b,c)=\pp{1_\ell\ot a&0\\ 0&1_p\ot c}.\eee The possible
complex conjugations in the representation will not be important in this
diagram. The mass matrix is
\bb \mm=\pp{M_1\ot 1_k&0\\ M_2\ot 1_k&0\\ 0&{M_3}^*\ot 1_p},\qq
M_1\in M_{k\times\ell}(\cc),\qq  M_2,M_3\in M_{p\times\ell}(\cc).\eee
The fluctuations are
\bb
\Mf_1  &=&\sum_j r_j\,u _jM_1 v_j^{-1},\qq u_j\in U(\Aa_1),\qq v_j\in
U(\Aa_2),\cr
\Mf_2  &=&\sum_j r_j\,w _jM_2 v_j^{-1},\qq w_j\in U(\Aa_3),\cr
\Mf_3  &=&\sum_j r_j\,w _jM_3 v_j^{-1},\eee and the action is, with
$C_i:={\Mf_i}^* \, \Mf_i$
\bb V(C_1,C_2,C_3)=4k \, [\lambda\, {\rm tr} (C_1+C_2)^2  -
{\textstyle\frac{1}{2}} \mu ^2\, {\rm tr} (C_1+C_2)] + 4p \, [\lambda\, {\rm tr}
(C_3)^2- {\textstyle\frac{1}{2}} \mu ^2 \,{\rm tr} (C_3)].
\eee Requiring at most one zero eigenvalue (up to a possible colour
degeneracy)  implies $k=1$, $\ell=p+1$ or $k=1$, $\ell=p$. In both cases
$\Mf_1$ and $\Mf_2$ vary independently. The colour group consists of the
$u$s and $w$s. As they are spontaneously broken we must have $k=p=1$,
leaving $\ell$ and $\ell=2$.
 In the commutative case with
no complex conjugations in the representation,
$\Mf_3=\beta \, \Mf_2$ for a complex constant $\beta \not= 0$. If
$|\beta |\ge 1$ the minimum is given by
$$|\Mfm_1|^2=
\frac{(1-|\beta |^{-2})\mu^2}{4\lambda}, \  |\Mfm_2|^2=
\frac{|\beta |^{-2}\mu^2}{4\lambda}, \ {\rm and} \ 
|\Mfm_3|^2=\frac{\mu^2}{4\lambda}.$$ Its mass spectrum
$\{0,\frac{\mu}{\sqrt{4\lambda}},\frac{\mu}{\sqrt{4\lambda}}\}$
 is dynamically degenerate. If
$|\beta |\le 1$ the minimum is given by
$$|\Mfm_1|^2= 0, \  |\Mfm_2|^2=\frac{1+|\beta |^2}{1+|\beta |^4}\
\frac{\mu^2}{4\lambda}, \ {\rm and} \  |\Mfm_3|^2=
\frac{|\beta |^2(1+|\beta |^2)}{1+|\beta |^4}\
\frac{\mu^2}{4\lambda}.$$ The triple is degenerate because of the
invariant subspace in the kernel of the Dirac operator. With additional
complex conjugations in the representation,
$\Mfm_3$ may decouple from $\Mfm_2$ and the triple becomes
dynamically degenerate as in diagram 1.

In the noncommutative case $k=1$, $\ell=2$, $p=1$, $M_1$ and
$M_2$ must be linearly independent. To get a nondegenerate triple for any
choice of $\Aa_1$ and $\Aa_3$ $=$ $\rr$ or $\cc$ and
$\Aa_2$ $=$ $M_2(\rr)$, $M_2(\cc)$ or $\hhh$ it is sufficient to take the
example: $M_1=(m_1,0),\ M_2=(0,m_2),\ M_3=\beta M_2$. The minimum is
given by
$\Mfm_1 \,{\Mfm_2}^*=0$ and
\bb |\Mfm_1|^2=\,\frac{\mu^2}{4\lambda}\, ,\qq
|\Mfm_2|^2=\,\frac{1+|\beta |^2}{1+|\beta  |^4}\,
\,\frac{\mu^2}{4\lambda}\, ,\qq |\Mfm_3|^2=|\beta  |^2\,\frac{1+|\beta  
|^2}{1+|\beta |^4}\,
\,\frac{\mu^2}{4\lambda}\, ,\eee dynamically non-degenerate for $\beta
\not= 0$ and $|\beta |\not= 1$.

\noindent Let us note a new phenomenon: the three eigenvalues of a
minimum $\ddfm$ are tied together by a {\it mass relation}.  Its origin is
clear, when we add more and more irreducible components to the Hilbert
space the number of possible fluctuations does not change, the number of
components in the Dirac operator increases. This phenomenon does not
only occur for the particular choice of the mass matrices $M_1$, $M_2$,
$M_3$ above, but is generic for diagram 8. The little groups are same as the
last six in (\ref{little}).

\vspace{1\baselineskip}
{\bf Diagram 9} yields the representations
\bb
\rho _L (a,b,c) =\pp{a\ot 1_k&0&0\\ 0&a\ot 1_p&0\\ 0& 0&b\ot 1_p}, \,
\rho _R (a,b,c) =\pp{b\ot 1_k&0&0\\ 0&a\ot 1_\ell&0\\
 0&0&c\ot 1_p},\cr \cr \cr
\rho _L^c (a,b,c) =\pp{1_k\ot a&0&0\\ 0&1_k\ot c&0\\ 0& 0&1_\ell\ot c}, \, 
\rho _R^c (a,b,c) =\pp{1_\ell\ot a&0&0\\ 0&1_k\ot b&0\\ 0& 0&1_p\ot c},
\eee with possible complex conjugations. The mass matrix is
\bb \mm=\pp{M_1\ot 1_k&0&0\\ 0&1_k\ot {M_2}^*&0\\ 0&0&M_3\ot
1_p},\, M_1\in M_{k\times\ell}(\cc),\,  M_2, M_3\in M_{\ell\times
p}(\cc).\eee
 The fluctuations are
\bb
\Mf_1  &=&\sum_j r_j\,u _jM_1 v_j^{-1},\qq u_j\in U(\Aa_1),\qq v_j\in
U(\Aa_2),\cr
\Mf_2  &=&\sum_j r_j\,v _jM_2 w_j^{-1},\qq w_j\in U(\Aa_3),\cr
\Mf_3  &=&\sum_j r_j\,v _jM_3 w_j^{-1},\eee and the action
$V(C_1,C_2,C_3)$ is equal to 
\bb 4k \, [\lambda\, {\rm tr} (C_1)^2 -  {\textstyle\frac{1}{2}} \mu ^2\,
{\rm tr} (C_1)] +4k \, [\lambda\, {\rm tr} (C_2)^2- {\textstyle\frac{1}{2}} \mu
^2 \,{\rm tr} (C_2)] +4p \, [\lambda\, {\rm tr} (C_3)^2- {\textstyle\frac{1}{2}}
\mu ^2 \,{\rm tr} (C_3)].
\eee 
Counting neutrinos and imposing broken colour to be commutative leaves
only one case, $k=\ell=p=1$. We choose $M_3=\beta 
M_2$ and get the minima at
\bb |\Mfm_1|^2=\,\frac{\mu^2}{4\lambda}\, ,\qq
|\Mfm_2|^2=\,\frac{1+|\beta |^2}{1+|\beta  |^4}\,
\,\frac{\mu^2}{4\lambda}\, ,\qq |\Mfm_3|^2=|\beta  |^2\,\frac{1+|\beta  
|^2}{1+|\beta |^4}\,
\,\frac{\mu^2}{4\lambda}\, ,\eee so it is dynamically non-degenerate for
$\beta \neq 0$ and $|\beta |\neq 1$.
\vspace{1\baselineskip}

{\bf Diagram 10} yields the representations
\bb \rho _L (a,b,c) =\pp{a\ot 1_k&0&0\\ 0&a\ot 1_k&0\\ 0& 0&b\ot 1_p},&&
\rho _R(a,b,c) =\pp{b\ot 1_k&0\\ 0&c\ot 1_p},\cr \cr \cr
\rho _L^c(a,b,c) =\pp{1_k\ot a&0&0\\ 0&1_k\ot a&0\\ 0& 0&1_\ell\ot c},&&
\rho _R^c(a,b,c) =\pp{1_\ell\ot a&0\\ 0&1_p\ot c},\eee with possible
complex conjugations. The mass matrix is
\bb \mm=\pp{M_1\ot 1_k&0\\ M_2\ot 1_k&0\\ 0&M_3\ot 1_p},\qq M_1,
M_2\in M_{k\times\ell}(\cc),\qq  M_3\in M_{\ell\times p}(\cc).\eee
 The fluctuations are
\bb
\Mf_1  &=&\sum_j r_j\,u _jM_1 v_j^{-1},\qq u_j\in U(\Aa_1),\qq v_j\in
U(\Aa_2),\cr
\Mf_2  &=&\sum_j r_j\,u _jM_2 v_j^{-1},\cr
\Mf_3  &=&\sum_j r_j\,v _jM_3 w_j^{-1},\qq w_j\in U(\Aa_3),
\eee and the action is
\bb V(C_1,C_2,C_3)=4k \, [\lambda\, {\rm tr} (C_1+C_2)^2- 
{\textstyle\frac{1}{2}} \mu ^2\, {\rm tr} (C_1+C_2)] + 4p \, [\lambda\, {\rm tr}
(C_3)^2- {\textstyle\frac{1}{2}} \mu ^2 \,{\rm tr} (C_3)].\eee Neutrino
counting implies
$k=1$, $\ell=2$, $p=1$ or $k=\ell=p=1$.
 In both cases $\Mf_3$ varies independently of $\Mf_1$ and
$\Mf_2$.

The commutative case has always $|\Mfm_1|^2 +|\Mfm
_2|^2=|\Mfm_3|^2=\mu ^2{(4\lambda)}^{-1}$ and is
 dynamically degenerate.

In the noncommutative case $k=1$, $\ell=2$, $p=1$, $M_1$ and
$M_2$ must be linearly independent. For
$\Aa_1\op\Aa_2=\rr\op M_2(\cc)$, $\cc\op M_2(\rr)$ and
$\cc\op M_2(\cc)$, lemma A.1 decouples $\Mf_1$ and $\Mf_2$ and the
triples are dynamically degenerate, $\Mfm_1=(\mu (4
\lambda )^{-1/2},0)$, $\Mfm_2=(0,\mu (4
\lambda )^{-1/2})$, $\Mfm_3=(0,\mu (4
\lambda )^{-1/2})^T$. For
$\Aa_1\op\Aa_2=\rr\op \hhh$ and $\cc\op \hhh$ we choose
\bb
\rho _L (a,b,c) =\pp{a\ot 1_k&0&0\\ 0&\bar a\ot 1_k&0\\ 0& 0&b\ot
1_p},\qq M_1=(m_1,0),\qq M_2=(0,\alpha m_1).
\eee 
Its minimum has the non-degenerate spectrum
$\{ \frac{1+|\alpha |^2}{1+|\alpha |^4},\ \frac{|\alpha |^2(1+|\alpha
|^2}{1+|\alpha |^4},\ 1,\ 0\}$ in units of $\frac{\mu ^2}{4\lambda}$ with one
mass relation. Finally for $\Aa_1\op\Aa_2=\rr\op M_2(\rr)$ we choose
$M_1=(i,1)$ and
$M_2=(1,0)$ to obtain the non-degenerate spectrum $\{ \frac{2 \pm \sqrt
2}{3},\ 1,\ 0\}$ in units of $\frac{\mu ^2}{4\lambda}.$
\vspace{1\baselineskip}

{\bf Diagram 11} yields the representations
\bb \rho _L (a,b,c) =\pp{a\ot 1_k&0\\
 0&b\ot 1_p},&&
\rho _R(a,b,c) =\pp{b\ot 1_k&0&0\\ 0&b\ot 1_k&0\\
 0&0&c\ot 1_p},\cr \cr \cr
\rho _L^c(a,b,c) =\pp{1_k\ot a&0\\
 0&1_\ell\ot c},&&
\rho _R^c(a,b,c) =\pp{1_\ell\ot a&0&0\\ 0&1_\ell\ot a&0\\ 0&0&1_p\ot
c},\eee with possible complex conjugations. The mass matrix is
\bb \mm=\pp{M_1\ot 1_k& M_2\ot 1_k&0\\0& 0&M_3\ot 1_p},\qq M_1,
M_2\in M_{k\times\ell}(\cc),\qq  M_3\in M_{\ell\times p}(\cc).\eee
 The fluctuations are
\bb
\Mf_1  &=&\sum_j r_j\,u _jM_1 v_j^{-1},\qq u_j\in U(\Aa_1),\qq v_j\in
U(\Aa_2),\cr
\Mf_2  &=&\sum_j r_j\,u _jM_2 v_j^{-1},\cr
\Mf_3  &=&\sum_j r_j\,v _jM_3 w_j^{-1},\qq w_j\in U(\Aa_3),
\eee and the action is
\bb V(C_1,C_2,C_3)=4k \, [\lambda\, {\rm tr} (C_1+C_2)^2- 
{\textstyle\frac{1}{2}} \mu ^2\, {\rm tr} (C_1+C_2)] + 4p \,  [\lambda\, {\rm tr}
(C_3)^2-  {\textstyle\frac{1}{2}} \mu ^2 \,{\rm tr} (C_3)].
\eee Neutrino counting  and imposing broken colour to be commutative
leaves only one possibility:
 $k=\ell=p=1$, which  is treated as the case
$k=\ell=p=1$ of diagram 10 with the same conclusion, degeneracy.
\vspace{1\baselineskip}

{\bf Diagram 12} yields the representations
\bb \rho _L (a,b,c) =\pp{a\ot 1_k&0\\
 0&b\ot 1_p},&&
\rho _R (a,b,c) =\pp{b\ot 1_k&0&0\\ 0&a\ot 1_\ell&0\\ 0&0&c\ot 1_p},\cr
\cr \cr
\rho _L^c (a,b,c) =\pp{1_k\ot a&0\\ 0&1_\ell\ot c},&&
\rho _R^c (a,b,c) =\pp{1_\ell\ot a&0&0\\ 0&1_k\ot b&0\\ 0&0&1_\ell\ot
c},\eee with possible complex conjugations. The mass matrix is
\bb \mm=\pp{M_1\ot 1_k& 1_k\ot M_2&0\\0& 0&M_3\ot 1_p},\qq M_1,
M_2\in M_{k\times\ell}(\cc),\ M_3\in M_{\ell\times p}(\cc).\eee The
fluctuations are
\bb
\Mf_1  &=&\sum_j r_j\,u _jM_1 v_j^{-1},\qq u_j\in U(\Aa_1),\qq v_j\in
U(\Aa_2),\cr
\Mf_2  &=&\sum_j r_j\,u _jM_2 v_j^{-1},\cr
\Mf_3  &=&\sum_j r_j\,v _jM_3 w_j^{-1},\qq w_j\in U(\Aa_3).
\eee The action $V(C_1,C_2,C_3)$ equals
\bb & 4k \, [\lambda\, {\rm tr} (C_1)^2- {\textstyle\frac{1}{2}} \mu ^2\,
{\rm tr} (C_1)] +4k  \, [\lambda\, {\rm tr} (C_2)^2- {\textstyle\frac{1}{2}} \mu
^2 \,{\rm tr} (C_2)] \cr & \hspace{1.3cm} + \, 8\lambda\, {\rm tr}
(C_1)\,{\rm tr} (C_2)  + 4p \, [\lambda\, {\rm tr} (C_3)^2-
{\textstyle\frac{1}{2}} \mu ^2 \,{\rm tr} (C_3)].
\eee Since
$\Mf_3 $ decouples, $\ell$ and $p$ can at most differ by one. The neutrino 
count and imposing broken colour to be commutative  implies $k=\ell=p=1$.
Then
$M_2=\alpha \, M_1$ and we must distinguish two cases: $\Mf_2=\alpha \, 
\Mf_1$ or ${}\Mf_1$ and $\Mf_2$ independent. Both possibilities have a
dynamically degenerate minimum: 
$|\Mfm_1|^2+|\Mfm_2|^2=|\Mfm_3|^2=\frac{\mu ^2}{4\lambda}$. 
\vspace{1\baselineskip}

{\bf Diagram 13} has a {\it ladder form}, i.e. it consists of horizontal
arrows,  vertically aligned.  Its representations are
\bb \rho _L (a,b,c) =\pp{a\ot 1_k&0&0\\ 0&a\ot 1_k&0\\ 0& 0&a\ot 1_p},&&
\rho _R (a,b,c) =\pp{b\ot 1_k&0\\ 0&b\ot 1_p},\cr \cr \cr
\rho _L^c (a,b,c) =\pp{1_k\ot a&0&0\\ 0&1_k\ot a&0\\ 0& 0&1_k\ot c},&&
\rho _R^c (a,b,c) =\pp{1_\ell\ot a&0\\ 0&1_\ell\ot c},\eee with possible
complex conjugations here and there. The mass matrix is
\bb \mm=\pp{M_1\ot 1_k&0\\ M_2\ot 1_k&0\\ 0&M_3\ot 1_p},\qq M_1,
M_2,M_3\in M_{k\times\ell}(\cc).\eee The fluctuations are
\bb
\Mf_1  &=&\sum_j r_j\,u _jM_1 v_j^{-1},\qq u_j\in U(\Aa_1),\qq v_j\in
U(\Aa_2),\cr
\Mf_2  &=&\sum_j r_j\,u _jM_2 v_j^{-1},\cr
\Mf_3  &=&\sum_j r_j\,u _jM_3 v_j^{-1},
\eee and the action is
\bb V(C_1,C_2,C_3) = 4k \, [\lambda\, {\rm tr} (C_1+C_2)^2 - 
{\textstyle\frac{1}{2}} \mu ^2\, {\rm tr} (C_1+C_2)] + 4p \, [\lambda\, {\rm tr}
(C_3)^2 - {\textstyle\frac{1}{2}} \mu ^2 \,{\rm tr} (C_3)].
\eee The neutrino count  implies $k=1$, $\ell=2$ or $k=\ell=1$.

\noindent The case $k=\ell =1$ has the following possibilities:

{\bf 1:} $\rr\oplus\rr\oplus\ccc$, where $\ccc$ is any simple `colour'
algebra. All possible triples are degenerate in the sense that the Dirac
operator has an invariant subspace in its kernel.

{\bf 2:} $\rr\oplus\cc\oplus\ccc$, all possible triples are degenerate.

{\bf 3:} $\cc\oplus\rr\oplus\ccc\owns (a,b,c)$. The non-degenerate
triples have:
\bb\rho_L (a,b,c) = \pp{a&0&0\\ 0&\bar a&0\\ 0&0&\bar a\ot 1_p},\qq
\rho_R (a,b,c) =\pp{b&0\\ 0&b\ot 1_p}.\eee The fluctuations respect the
mass ratios:
$|M_1|:|M_2|:|M_3|= |\Mf_1|:|\Mf_2|:|\Mf_3|$. If $M_1$ and $M_2$ are
different from zero, the kernel of the Dirac operator, $\cc
(-\overline{M_2},\overline{M_1},0,0,0;$
$-M_2,M_1,0,0,0)^T,$ is not invariant under $a\in
\cc.$ If $|M_3|^2\not= |M_1|^2+|M_2|^2$, the triple is dynamically
non-degenerate.

{\bf 4:}  $\cc\oplus\cc\oplus\ccc\owns (a,b,c)$. The only not obviously
degenerate
 triples have the representations:
\bb
\rho_L (a,b,c) = \pp{a&0&0\\ 0&\bar a&0\\ 0&0&\bar a\ot 1_p},\qq
\rho_R (a,b,c) =\pp{b&0\\ 0&b\ot 1_p},\label{cc1}
\ee and
\bb
\rho_L (a,b,c) = \pp{a&0&0\\ 0&\bar a&0\\ 0&0&\bar a\ot 1_p},\qq
\rho_R (a,b,c) =\pp{b&0\\ 0&\bar b\ot 1_p}.\label{cc2}
\ee The fluctuations do not respect all mass ratios, in fact
$\Mf_1$ and $\Mf_2$ are independent variables and only the ratio
$M_2/M_3=\Mf_2/\Mf_3=:1/\beta$  is invariant under fluctuations via
(\ref{cc1}). If
$|\beta |\ge 1$ the minima are given by
$$ |\Mfm_1|^2=\frac{(1-|\beta |^{-2})\mu^2}{4\lambda}, \ Ê
 |\Mfm_2|^2=\frac{\mu^2}{|\beta |^2 \; 4\lambda}, \   {\rm and} \
|\Mfm_3|^2=\frac{\mu^2}{4\lambda}.
$$  Its mass spectrum
$\{0,\frac{\mu}{2\sqrt{\lambda}},\frac{\mu}{2\sqrt{\lambda}} \, { p}
\,{\rm times}\}$
 is dynamically degenerate. If
$|\beta |\leq 1$  the minima are given by
$$|\Mfm_1|^2=0, \  |\Mfm_2|^2=\frac{1+p|\beta |^2}{1+p|\beta |^4} \,
\frac{\mu^2}{4\lambda}, \   {\rm and} \  |\Mfm_3|^2=\frac{|\beta
|^2(1+p|\beta |^2)}{1+p|\beta |^4} \,\frac{\mu^2}{4\lambda}.$$  The triple is
degenerate since there is an invariant subspace in the kernel of the Dirac
operator. For the representation (\ref{cc2}) the computations are identical
to the case above, (\ref{cc1}), after permuting $M_1$ and
$M_2$ and after replacing $M_3$ by $\overline{M_3}$.

In the case $k=1,$ $\ell =2$,  the three submatrices $M_1,$ $M_2,$ $M_3$
are linearly dependent over $\cc$. If
$M_1$ is proportional to $M_2$ then both $M_3$ and $\pp{M_1\\ M_2}$
have a zero eigenvalue. Therefore $M_1$ and $M_2$ are linearly
independent over $\cc$ and $M_3=:\alpha M_1+\beta M_2$ with complex
coefficients $\alpha $ and $\beta $.

{\bf 5:} $\cc\oplus M_2(\cc)\oplus\ccc\owns (a,b,c)$. By lemma A.1,
$\Mf_1$ and
$\Mf_2$ vary independently in $\cc^2$. For the representations with
\bb\rho_L(a,b,c)= \pp{a&0&0\\ 0& a&0\\ 0&0& a\ot 1_p},\qq
\rho_R(a,b,c)=\pp{b&0\\ 0&b\ot 1_p},\label{realrep}\ee we get
$\Mf_3=:\alpha \Mf_1+\beta \Mf_2$. For instance for $\alpha = 0$ we get
$\Mfm_1 \, {\Mfm_2}^*=0$ as for diagram 8 and a similar mass relation,
\bb |\Mfm_1|^2=\,\frac{\mu^2}{4\lambda}\, ,\qq
|\Mfm_2|^2=\,\frac{1+p|\beta |^2}{1+p|\beta  |^4}\,
\,\frac{\mu^2}{4\lambda}\, ,\qq |\Mfm_3|^2=|\beta  |^2\,\frac{1+p|\beta  
|^2}{1+p|\beta |^4}\,
\,\frac{\mu^2}{4\lambda}\, .\label{massrel}\ee 
For general $\alpha$, we get the same relations with 
$|\beta|^2$ replaced by $|\alpha|^2 +|\beta|^2$. 
For the representations with
\bb\rho_L(a,b,c)= \pp{a&0&0\\ 0&\bar a&0\\ 0&0& \bar a\ot 1_p},\qq
\rho_R(a,b,c)=\pp{b&0\\ 0&b\ot 1_p},\eee and $\alpha  \not= 0$, all three
doublets, $\Mf_1,$ $\Mf_2$ and
$\Mf_3$ vary independently. The minima,
$|\Mfm_1|^2=|\Mfm_2|^2=|\Mfm_3|^2=\frac{\mu ^2}{4\lambda}$ produce a
dynamical degeneracy in this case. The other case, $\alpha  = 0$ is
dynamically non-degenerate with the same
 mass relation as above, equations (\ref{massrel}). All other
representations of this algebra are treated the same way and they either
have a mass relation or are  dynamically degenerate, which means a
particularly simple mass relation.

{\bf 6:} $\rr\oplus M_2(\cc)\oplus\ccc$. This case is identical to case {\bf
5} with representation (\ref{realrep}).

{\bf 7:}
$\cc\oplus\hhh\oplus\ccc\owns (a,b,c)$. For example, all triples with
\bb\rho_L(a,b,c)= \pp{a&0&0\\ 0&\bar a&0\\ 0&0&\bar  a\ot 1_p},\qq
\rho_R(a,b,c)=\pp{b&0\\ 0&b\ot 1_p},\eee
$M_1=(m_1,0)$, $M_2=(0,m_2)$ and $M_3=(0,m_3)$ have no mass relation.
Indeed like in case {\bf 3},  the mass ratios are stable under fluctuations:
$|M_1|:|M_2|:|M_3|= |\Mf_1|:|\Mf_2|:|\Mf_3|$. Other triples behave like in
case {\bf 5}.

{\bf 8:} $\rr\oplus\hhh\oplus\ccc$ has the same examples without mass
relations as in case {\bf 7}.

{\bf 9:}  $\cc\oplus  M_2(\rr)\oplus\ccc$. For representations
(\ref{realrep}) and $\alpha =0$, we get minima with mass relation
(\ref{massrel}).

{\bf 10:} $\rr\oplus M_2(\rr)\oplus\ccc$. Here we take a representation
(\ref{realrep}) with $M_1=(m_1,0)$, $M_2=(0,m_2)$ and $M_3=(0,\beta \,
m_2)$ and get again the mass relation (\ref{massrel}).

Up to different multiplicities, we have the same conclusion for the {\bf
diagrams 18, 17, 22.} After permuting $\Aa_1$ and $\Aa_2$ we also have
the same results for the other four ladders, {\bf diagrams 14, 19, 16, 21.}
\vspace{1\baselineskip}

{\bf Diagram 15} yields the representations
\bb \rho _L (a,b,c) =\pp{a\ot 1_k&0\\ 0&a\ot 1_p},&&
\rho _R (a,b,c) =\pp{b\ot 1_k&0&0\\ 0&a\ot 1_\ell&0\\ 0&0&b\ot 1_p},\cr
\cr \cr
\rho _L^c (a,b,c) =\pp{1_k\ot a&0\\ 0&1_k\ot c},&&
\rho _R^c (a,b,c) =\pp{1_\ell\ot a&0&0\\ 0&1_k\ot b&0\\ 0&0&1_\ell\ot
c},\eee with possible complex conjugations. The mass matrix is
\bb \mm=\pp{M_1\ot 1_k& 1_k\ot M_2&0\\0& 0&M_3\ot 1_p},\qq M_1,
M_2,M_3\in M_{k\times\ell}(\cc).\eee The fluctuations are
\bb
\Mf_1 &=&\sum_j r_j\,u _jM_1 v_j^{-1},\qq u_j\in U(\Aa_1),\qq v_j\in
U(\Aa_2),\cr
\Mf_2  &=&\sum_j r_j\,u _jM_2 v_j^{-1},\cr
\Mf_3  &=&\sum_j r_j\,u _jM_3 v_j^{-1}.
\eee Neutrino counting and imposing broken colour to be commutative
implies  $k=\ell=1$. This case with $\Aa_1$ and $\Aa_2$ = $\rr$
or $\cc$ is treated as in diagram 13, {\bf 3} yielding a non-degenerate
triple without mass relation. After replacing
$M_3^*$ by
$M_3$, {\bf diagram 20} has identical computations.
\vspace{1\baselineskip}

{\bf Diagram 23} yields the representations
\bb \rho _L (a,b,c) =\pp{a\ot 1_k&0&0\\ 0&b\ot 1_\ell&0\\ 0& 0&b\ot
1_p},&&
\rho _R (a,b,c) =\pp{b\ot 1_k&0\\ 0&a\ot 1_p},\cr \cr \cr
\rho _L^c (a,b,c) =\pp{1_k\ot a&0&0\\ 0&1_\ell\ot b&0\\ 0& 0&1_\ell\ot
c},&&
\rho _R^c (a,b,c) =\pp{1_p\ot a&0\\ 0&1_k\ot c},\eee with possible complex
conjugations. The mass matrix is
\bb \mm=\pp{M_1\ot 1_k&0\\ 1_\ell\ot M_2^*&0\\ 0&M^*_3\ot 1_p},\qq
M_1, M_2,M_3\in M_{k\times\ell}(\cc).\eee The fluctuations are
\bb
\Mf_1  &=&\sum_j r_j\,u _jM_1 v_j^{-1},\qq u_j\in U(\Aa_1),\qq v_j\in
U(\Aa_2),\cr
\Mf_2  &=&\sum_j r_j\,u _jM_2 v_j^{-1},\cr
\Mf_3  &=&\sum_j r_j\,u _jM_3 v_j^{-1}.
\eee Neutrino counting   implies $k=\ell=1$. This model with $\Aa_1$ and
$\Aa_2$ =
$\rr$ or $\cc$ is treated as in diagram 13, {\bf 3} yielding a
non-degenerate triple without mass relation. After replacing $M_3^*$ by
$M_3$, {\bf diagram 24} has identical computations.

We must now extend our analysis to include the possibility of complex
conjugate representations. As in the case of two algebras, one shows that
any diagram containing a connected component consisting of only letter
unchanging arrows leads to degenerate spectra. Therefore within the
class of irreducible and dynamically non-degenerate triples the leitmotiv
of a Krajewski diagram is still carried by its letter changing arrows. For
three algebras there are two additional diagrams, {\bf diagrams 25, 26},
figure 7, that  involve only letter changing arrows.   Their contracted
diagrams resemble diagrams 4 and 9 of figure 6 except  for the change of
chirality in one arrow and the computation of their triples is similar.

Note that without the presence of conjugate representations, this change
violates the condition  that nonvanishing entries of the multiplicity
matrix and its transposed must have same signs. Figure 8 lists the
contractions of all irreducible diagrams whose letter unchanging arrows
are connected to at least one letter changing arrow. The blow up of the
new symbols is given in Figure 9.

All triples attached to the eleven diagrams of Figure 8 are degenerate or
dynamically degenerate:
\vspace{1\baselineskip}

{\bf Diagram 27} with the first blow up yields
\bb
\rho _L (a,b,c) =\pp{c\ot 1_k&0\\ 0&b\ot 1_p},\qq
\rho _R (a,b,c) =\pp{b\ot 1_k&0&0\\ 0&\bar c\ot 1_k&0\\
 0&0&a\ot 1_p},
\eee
\vspace{-0.5cm}
\bb
\mm=\pp{M_1\ot 1_k&M_2\ot 1_k&0\\ 0&0&M_3\ot 1_p},
\eee
$ M_1\in M_{p\times\ell}(\cc),\  M_2\in M_{p\times p}(\cc),\  M_3\in
M_{\ell\times k}(\cc).$ Counting neutrinos leads to $k=\ell=1$. To get a
handle on $p$ we repeat the overkill from section 4. In the worst case
$M_2$ and consequently $\Mf_2$ is skewsymmetric and $p$ is odd. By
fluctuations we can obtain
\bb
\Mf_2=\pp{\pp{0&-1\\ 1&0}&&&\\ &\pp{0&-1\\ 1&0}&&\\ &&\ddots&\\
&&&0}=:\pp{A&0\\ 0&0}.
\eee Now $\Mf_1$ fluctuates independently and we may obtain for its
transpose 
${\Mf_1}^T=(0,...,0,1)$. We get
$(\Mf_1,\Mf_2){(\Mf_1,\Mf_2)}^*=1_p$ and the minimum of the action is
dynamically degenerate if $p\ge 1$. The commutative case $k=\ell=p=1$ is
obviously degenerate. For the second blow up the computations are similar
with $\Mf_2$ now proportional to $1_p$ from the start. 
\vspace{1\baselineskip}

{\bf Diagram 28} is treated as diagram 27. 
\vspace{1\baselineskip}

{\bf Diagram 29} has $k=\ell=p=1$ by neutrino count and admits only
degenerate triples. 
\vspace{1\baselineskip}

{\bf Diagrams 30, 31, 32, 34, 35, 36} have $k=\ell =1$ by neutrino count. As
in the commutative case of diagram 8, all their triples are degenerate. 
\vspace{1\baselineskip}

{\bf Diagram 33} with the first of the six possible blow ups yields:
\bb
\rho _L (a,b,c) &=&\pp{a\ot 1_k&0&0\\ 0&a\ot 1_p&0\\ 0&0&\bar b\ot
1_p}, \nonumber \\[2mm]
\rho _R (a,b,c) &=&\pp{b\ot 1_k&0&0&0\\ 0&b\ot 1_p&0&0\\ 0&0& b\ot
1_p&0\\0&0&0&b\ot 1_p}, \nonumber \\[2mm]
\rho ^c_L (a,b,c) &=&\pp{1_k\ot a&0&0\\ 0&1_k\ot c&0\\ 0&0&1_\ell\ot
c}, \nonumber \\[2mm]
\rho^c_R (a,b,c) &=&\pp{1_\ell\ot a&0&0&0\\ 0&1_\ell\ot c&0&0\\ 0&0&
1_\ell\ot c&0\\0&0&0&1_\ell\ot c}, \nonumber \\[2mm]
\mm&=&\pp{M_1\ot 1_k&0&0&0\\ 0&M_2\ot 1_p&0&0\\0&M_3\ot 1_p&
M_4\ot 1_p&M_5\ot 1_p},\eee
$M_1,M_2 \in M_{k\times\ell}(\cc),\  M_3,M_4,M_5\in
M_{\ell\times\ell}(\cc).$ Counting neutrinos and imposing broken colour
to be commutative gives $k=\ell=1$. This case is
degenerate: the kernel of the Dirac operator contains the invariant subspace
with elements
  $(0,0,0,0,0,M_5 v,-M_4 v;$ $0,0,0,0,0,\overline{M_5 w},\overline{-M_4
w})^T$, $v,w\in\cc^p$.

With the second blow up diagram 33 yields
\bb
\rho _L (a,b,c) &=&\pp{a\ot 1_k&0&0\\ 0&a\ot 1_p&0\\ 0&0&\bar b\ot
1_p}, \nonumber \\[2mm]
\rho _R (a,b,c) &=&\pp{b\ot 1_k&0&0&0\\ 0&b\ot 1_p&0&0\\ 0&0& b\ot
1_p&0\\0&0&0&\bar b\ot 1_p}, \nonumber \\[2mm]
\rho ^c_L (a,b,c) &=&\pp{1_k\ot a&0&0\\ 0&1_k\ot c&0\\ 0&0&1_\ell\ot
c}, \nonumber \\[2mm]
\rho^c_R (a,b,c) &=&\pp{1_\ell\ot a&0&0&0\\ 0&1_\ell\ot c&0&0\\ 0&0&
1_\ell\ot c&0\\0&0&0&1_\ell\ot\bar c}, \nonumber \\[2mm]
\mm&=&\pp{M_1\ot 1_k&0&0&0\\ 0&M_2\ot 1_p&0&0\\0&M_3\ot 1_p&
M_4\ot 1_p&1_k\ot M_5},\eee
$M_1,M_2 \in M_{k\times\ell}(\cc),\  M_3,M_4\in
M_{\ell\times\ell}(\cc),\ M_5\in
M_{p\times p}(\cc).$ Counting neutrinos and imposing broken colour
to be commutative gives $k=\ell=p=1$. In the notations of Corollary 11.4, the 
action reads
\bb V(a,b,c)&=&4\lambda\,
[|M_1ab|^4+|M_2ab|^4+2|M_2ab|^2|M_3b^2|^2+
(|M_3b^2|^2+|M_4b^2|^2+|M_5c^2|^2)^2]\nonumber\\
&&-2\mu ^2\,[|M_1ab|^2+|M_2ab|^2+(|M_3b^2|^2+|M_4b^2|^2+|M_5c^2|^2)].\eee 
Its minimum is degenerate, $\hat b=0$.

 The other blow ups as well as {\bf
diagram 37} lead to the same conclusion: degeneracy.\\ This completes the
proof of the theorem.\qed

We summarize the possible algebras  with $N=3$ and the corresponding
Krajewski diagrams of all their irreducible, dynamically non-degenerate
triplets in a table:

 \begin{center}
\begin{tabular}{|c|c|}
\hline &\\
 Algebra  & Diagrams\\[1ex]
\hline &\\
$\cc\op\rr\op(\ccc=\bf 1)$&14,16,19,21\\[1ex]
$\cc\op\rr\op\ccc$&13,17,18,22\\[1ex]
${\bf 1}\op{\bf 1}\op\ccc$&15,20,23,24\\[1ex]
${\bf 2}\op{\bf 1}\op(\ccc=\bf 1)$&14,16,19,21\\[1ex]
${\bf 2}\op{\bf 1}\op\ccc$&13,17,18,22\\[1ex]
${\cc}\op{\cc}\op {\bf 1}$&26\\[1ex]
${\bf 1}\op{\bf 1}\op {\bf 1}$&9\\[1ex]
$M_2(\rr)\op\rr\op {\bf 1}$&10\\[1ex]
$\hhh\op{\bf 1}\op {\bf 1}$&10\\[1ex]
${\bf 2}\op{\bf 1}\op {\bf 1}$&8\\[1ex]
\hline
\end{tabular}
\end{center} 
Note that relaxing the hypothesis of unbroken noncommutative colour
does not add any algebra to the list with $N=1$ and 2. It adds only few
algebras to the list with $N=3$ coming from diagrams 8, 9 and 11. We were
unable to treat some of their triples, in particular quaternionic ones.

\subsection{The standard model of electro-weak and strong forces} Let us
close this section by remarking that diagram 17 of Figure 6 with flipped
chirality and with algebra
$\Aa=\cc\op\hhh\op M_3(\cc)\owns (b,a,c)$, with representation
\bb
\rho _L (a,b) =\pp{a\ot1_3&0\\ 0& a},\qq
\rho _R (a,b) =\pp{b1_3&0&0\\ 0&\bar b1_3&0\\ 0&0&\bar b},
\eee
\vspace{-0.5cm}
\bb 
\rho _L^c (a,b) =\pp{1_2\ot c&0\\ 0&\bar b1_2},\qq
\rho _R^c (a,b) =\pp{c&0&0\\ 0&c&0\\ 0&0&\bar b},
\eee 
and with mass matrix
\bb
\mm= \pp{\pp{m_u\\ 0}\ot 1_3&\pp{0\\ m_d}\ot 1_3&0\\
0&0&\pp{0\\ m_e}}
\eee 
produces the standard model of electro-magnetic,
weak and strong forces with one generation of quarks and leptons,  $u$,
$d$, $ \nu $ and $e$. The neutrino is a massless Weyl spinor.
The intersection form written with respect to the basis of projectors
\bb p_1=(0,1_2,0),\qq p_2=(1,0,0),\qq p_3=\left( 0,0,\pp{1&0&0\\
0&0&0\\0&0&0}\right) \eee
is
\bb \cap=-2\pp{0&1&1\\1&-1&-1\\1&-1&0},\eee
and non-degenerate.
 The colour group $U(3)$ is unbroken and its
representations on corresponding left- and right-handed fermions are
identical. In physicists' language this means that gluons are massless and
couple vectorially. Further details on the standard model as an almost
commutative geometry can be found in \cite{rep,bridge}.

\section{Beyond irreducible triples \label{reduc}}

For the standard model, allowing reducible triples has two important
physical consequences:

i) Suppose we want to render the neutrino massive.  Majorana
masses are incompatible with the axiom that the Dirac operator
anticommutes with the chirality. Therefore we must increase the Hilbert
space by adding a right-handed neutrino. Then the triple becomes
reducible, but worse Poincar\'e duality breaks down: the intersection form
becomes 
\bb \cap=-2\pp{0&1&1\\1&-2&-1\\1&-1&0},\eee 
degenerate.

ii) We may add more generations of quarks and leptons. Then the
Cabibbo-Kobayashi-Maskawa matrix makes its appearance. Now we may add
 right-handed neutrinos in some but not in all generations and give Dirac
masses to the corresponding neutrinos without violating Poincar\'e duality. 

So far we have no clue to why the standard model comes with three
colours, with three generations of quarks and with three generations of
leptons. Note however that anomaly cancellations \cite{anom} imply further
constraints that are satisfied with three colours and with a number of quark
generations equal to the number of lepton generations.

\subsection{A reducible triple with non-degenerate spectrum}
\label{exemple}

The criterion of dynamical degeneracy loses its meaning in presence of
reducible triples as illustrated by the following example:
$\Aa=M_3(\cc)\op M_2(\cc)\op\cc\owns (a,b,c)$,
\bb \rho (a,b,c)=\pp{ a\ot 1_2&0&0&0\\ 0&b\ot 1_3&0&0\\ 0&0&\bar
c1_3\ot 1_2&0\\ 0&0&0&\bar c1_2\ot 1_3},\
\eee
\bb \mm=m\pp{
\sqrt{\frac{5}{17} }\pp{1&0&0\\ 0&2&0}&0\\ 0&\,\frac{1}{\sqrt
5}\,\pp{1&0&-2\\ 0&2&0}\\
\pp{0&0&1\\ 0&0&0}&\,\frac{2}{\sqrt 5}\, \pp{ 0&0&0\\ 1&0&0}}.\eee The
fluctuations of the Dirac operator generate an 18 dimensional complex
vector space in which the spectral action $V(\ddf)$ has to be minimized.
We used a steepest descend method of Mathematica for this task and found
an absolute minimum at
$\ddfm=\dd$ with $m=\frac{\mu}{\sqrt {4\lambda}}$ and $ V(\ddfm)=-
{\textstyle\frac{431}{340}} {\textstyle\frac{\mu ^4}{\lambda }} $. The
spectrum of this minimum $\ddfm$ is non-degenerate: in units of
$\frac{\mu}{\sqrt {4\lambda}} $ we have
$\{1,\ \frac{4}{5},\ \frac{5}{17},\ \frac{20}{17},\ 
\frac{9\pm\sqrt{17}}{10}\}$. All six values of course appear twice with a
positive and twice with a negative sign in the spectrum of $\ddfm$. The
little group of this minimum is $G_\ell= U(1)\times U(1)\subset U(3)\times
U(2)\times U(1)$ with generic element $(e^{i\alpha }1_3,e^{i\alpha
}1_2,e^{i\gamma  }).$ The spectrum of the minimum appears completely
rigid, i.e. there are mass relations.

\section{Conclusion}

Suppose we want to apply  conventional, perturbative quantum field
theory to the Yang-Mills-Higgs models coming from almost commutative
geometries. Then after renormalization, fermion masses are functions of
energy and the colour degeneracy is compatible with this energy
dependence only if all noncommutative colour groups are unbroken.
 Furthermore the renormalization of fermion masses is
incompatible with mass relations, in particular with the completely rigid
reducible spectral triple of section
\ref{reduc}. In irreducible spectral triples, mass relations other than
degeneracies only appear starting with
$N=3$. All triples without such mass relations come from ladder diagrams
and have algebras
$\bf 1\op 1 \op
\ccc$ or $\hhh\op \bf 1\op\ccc$.

Let us suppose  that also for $N\ge 4$ the irreducible triples without mass
relations have
 contracted multiplicity matrices of ladder type:
\bb
\hat \mu =\pp{\alpha &\beta &0&0\\
\gamma &\delta &0&0\\
\rho &\sigma &0&0\\
\theta  &\xi &0&0},\qq N=4.
\eee

Then det$(\hat\mu +\hat\mu ^T)=(\rho \xi -\sigma \theta )^2$ and
irreducibility implies $\alpha =\beta =\gamma =\delta =0$. For $N\ge 5$, all
contracted multiplicity matrices $\hat\mu $ of ladder type
 have det$(\hat\mu +\hat\mu ^T)=0$ leading us to the

\begin{conj}
 The sum of $N$ simple algebras,
$\Aa=\Aa_1\op\Aa_2\op ...\op\Aa_N$ admits a finite, real, $S^0$-real,
irreducible and dynamically non-degenerate spectral triple free of mass
relations if and only if it is in the list, up to a reordering of the summands:
\begin{center}
\begin{tabular}{|c|r|c|r|c|}
\hline &&&&\\
 $N=1$&$N=2\qq$&$N=3$&$N=4$\qq\qq\qq&$N\ge 5$\\[1ex]
\hline &&&&\\
&$\rr\op\rr$&$\rr\op\rr\op\ccc$&$\rr\op\rr\op\ccc_1\op
\ccc_2$&\\[1ex]
&$\rr\op\cc$&$\rr\op\cc\op\ccc$&$\rr\op\cc\op\ccc_1\op
\ccc_2$&\\[1ex]
&$\cc\op\cc$&$\cc\op\cc\op\ccc$&$\cc\op\cc\op\ccc_1\op
\ccc_2$&\\[1ex] &$M_2(\rr)\op\rr$&& &\\[1ex] {\rm
void}&$M_2(\rr)\op\cc$&&&{\rm void}\\[1ex]
&$M_2(\cc)\op\rr$&&&\\[1ex] &$M_2(\cc)\op\cc$&&&\\[1ex]
&$\hhh\op\rr$&$\hhh\op\rr\op\ccc$&
$\hhh\op\rr\op\ccc_1\op
\ccc_2$&\\[1ex] &$\hhh\op\cc$&$\hhh\op\cc\op\ccc$&
$\hhh\op\cc\op\ccc_1\op
\ccc_2$&\\[1ex]
\hline
\end{tabular}
\end{center} Here $\ccc$, $\ccc_1$ and $\ccc_2$ are three arbitrary
simple algebras. The colour algebras $\ccc$ for $N=3$ and
$\ccc_1\op
\ccc_2$ for $N=4$ have two  constraints:

i) Their representations on corresponding left- and right-handed
subspaces of $\hh$ are identical (up to possibly different multiplicities). 

ii) The Dirac operator $\dd $  is invariant under $U(\ccc)$ or
$U(\ccc_1\op\ccc_2)$,
\bb \rho (1,1,w) \,\dd \, \rho (1,1,w)^{-1}=\dd,\qq {\rm for \ all}\ w\in
U(\ccc)\ {\rm or}\ U(\ccc_1\op
\ccc_2).\eee This implies that the unitaries of ${}\ \ccc$ or $\ccc_1\op
\ccc_2$ do not participate in the fluctuations and are therefore unbroken,
i.e. elements of the little group.
\end{conj}

We must admit that our brute force proof by exhaustion is not suitable for
$N = 4$ and it seems already a formidable task to write down the list of all
contracted irreducible diagrams.

Besides renormalizability, there are two other important items on the
physicist's shopping list, which will further constrain the model building
kit:

i) The electric charge of a massless particle must be zero.

ii) The representation of the little group on the Hilbert space of fermions
must be complex.

Recall that a unitary representation is called real if it is equal to its
complex conjugate and pseudo-real if it is unitarily equivalent to its
complex conjugate. Otherwise the representation is complex. For example
the fundamental representation of $SU(2)$ is pseudo-real. An irreducible,
unitary representation of $U(1)$ is complex if and only if its charge is
non-zero.

Before we can examine these two criteria in the irreducible context, we
must compute the minimal central extensions \cite{min,Zou} that allow the
lift of algebra automorphisms to the Hilbert space of fermions to have at
most a finite number of values. This calculation is under way.

\vskip1cm
\noindent
{\bf Acknowledgements:} One of us (T.S.) thanks Harald Grosse for his hospitality
at the Erwin Schr\"odinger Institute where part of this work was done. We
gratefully acknowledge a fellowship of the Friedrich-Ebert-Stiftung for C.S.

\section{Appendix}

Since this paper deals with matrices,  let us briefly recall two standard
results,  one on the singular value decomposition of rectangular matrices
and the  second on the standard form of skewsymmetric matrices.

We write
$O(n):=U(M_n(\rr)),\ U(n):=U(M_n(\cc))$ and $USp(n):=U(M_n(\hhh))$.

\begin{lem}
i) Let $M \in M_{n \times m}(\cc)$. Then there exist $U\in U(n), V\in
U(m)$  such that $M=U \, D \, V$ where $D \in M_{n \times m}(\rr)$
satisfies $D_{ij}=0$ for 
$i \neq j$ and $D_{11} \geq D_{22} \geq D_{kk} >D_{k+1,k+1}= \cdots = D_{qq}=0$ 
where $k={\rm rank}(M)$ and $q={\rm min}(n,m)$. The $D_{ii}^2$ are 
the eigenvalues of $M^*M$, the columns of $U$ (resp. $V$) 
are the eigenvectors of $MM^*$  (resp. $M^*M$) arranged in the same 
order as the eigenvalues $D_{ii}^2$. In particular, when $M \in M_{n \times m}(\rr)$, 
we may assume $U \in O(n), V \in O(m)$ (\cite{Horn} 7.3.5).

ii) Let $M \in M_n(\cc)$ be a skewsymmetric matrix. Then, there exists $U \in U(n)$ 
such that 
\bb \label{antisym}
 UMU^T=\left( \bigoplus_{i=1}^p m_i x\right)  \oplus0 \cdots\oplus0 \qq {\rm
where}\qq x=\pp{0&1\\ -1&0},
\qq m_i \in \cc^* 
\ee 
and the numbers of zeros equals $n-2p$ (\cite{Horn} 4.4, Problem 26). 
\end{lem}

For our purpose, $M$ is the {\it complex} fermionic mass matrix. Then, in $i)$, the
diagonal elements $D_{jj}$ are the Dirac masses and the unitaries $U$ and
$V$ are related to the Cabibbo-Kobayashi-Maskawa matrix.

\begin{defn} Let $M \in M_{n \times m}(\cc)$ and $$f \in
\mathcal{F}_{\kk,\kk'}:=
 \{(r_j,u_j,v_j)_{j \in J} \in \rr \times U(M_n(\kk)) \times U(M_m(\kk'))
\mid J{\rm \ finite} \}$$ where 
$\kk,\kk'$ are $\rr, \, \cc$ or $\hhh$. 
 The {\it fluctuation} of $M$
is defined by
$$\Mf:= \sum_j r_j  \, u_j M v_j.$$ In the case that $\cc$ and $\hhh$ are
involved, we assume of course that $\cc \subset \hhh$.

Note that for a given $M$, 
$$\{\Mf \mid f \in \mathcal{F}_{\rr,\cc} \} = \{\Mf \mid f \in
\mathcal{F}_{\cc,\rr} \} = 
\{\Mf \mid f \in \mathcal{F}_{\cc,\cc} \}.$$
\end{defn}

\begin{lem} \label{genere} Let ${\rm Span}_{\rr}(E)$ be the real vector space
spanned by the set $E$. Then,

i) ${\rm Span}_{\rr}(O(n))=M_n(\rr)$.

ii) ${\rm Span}_{\rr}(U(n))=M_n(\cc)$.

iii) ${\rm Span}_{\rr}(USp(n))=M_n(\hhh)$.
\end{lem}

\begin{proof}  i) It is sufficient to prove that any $a=\pm a^T \in
M_n(\rr)$ is in ${\rm Span}_{\rr}(O(n))$.

When $a=a^T$, there exists $v \in O(n)$ such that $a=v\,d\,v^T$ where
$d$  is a real diagonal matrix. Since 
$$d=\sum_{i=1}^n {\textstyle\frac{d_{ii}}{2}} \, (2
p_i-1_n)+{\textstyle\frac{d_{ii}}{2}} \, 1_n,$$ where $p_i$ is the
projection on the i-th vector basis, so  
$d$ is in ${\rm Span}_{\rr}(O(n))$ and so is $a$.

When $a=-a^T$, there exist $v \in O(n)$ and a family $r_k \in \rr$, $k
\leq \frac{n}{2}$ such that
$$ a= v \, {\rm diag}(0,\cdots , 0 , r_1b , \cdots , r_kb) \, v^T$$ where
$b=\pp{0&1\\-1&0}$. Thus for $r=\sum_i r_i$, 
\bb v^Tav=b_1 \, {\rm diag} (1,\cdots,1,b,1_2,\cdots,1_2)+r_2 \, {\rm diag}
(1,\cdots,1,1_2,b,1_2\cdots,1_2) +\cdots + \cr +r_k \,
(1,\cdots,1,1_2,\cdots,1_2,b) - {\rm diag}
(r,\cdots,r,(r-r_1)1_2,\cdots,(r-r_k)1_2)
\eee is a real linear combination of matrices in $O(n)$ by i).  So $a \in
{\rm Span}_{\rr}(O(n))$.

ii) This follows by i) since $O(n)$ and $iO(n)$ are included in
$U(n)$.

iii) Let ${1,e_1,e_2,e_3}$ be the canonical basis of $\hhh$ such that 
$e_ie_j=\delta_{ij}1-\epsilon_{ijk}e_k$ and $1e_i=e_i1=e_i$. Since
$M_n(\hhh)$ is an 
$\hhh$-vector space,
$M_n(\hhh)=1M_n(\rr)+e_1M_n(\rr)+e_2M_n(\rr)+e_3M_n(\rr)$ and 
the result follows from $e_iO(n) \subset USp(n)$.  
\end{proof}

\begin{cly} $\{ \Mf \mid f \in \mathcal{F}_{\kk,\kk'} \} =  
\{\sum_i a_i \, M\, b_i \mid \, a_i \in M_{n}(\kk),b_i \in  M_{m}(\kk') \}.$
\end{cly}

\begin{rem} If $\kk=\kk'$, then for any $0 \neq M \in M_{n \times
m}(\kk)$, 
$$\{ \Mf \mid f \in \mathcal{F}_{\kk,\kk} \} = M_{n \times m}(\kk).$$
Nevertheless, we have a priori
$$\{ \Mf \mid f \in \mathcal{F}_{\kk,\kk'} \} \  \hbox{$\subset$
\hspace{-0.37cm}/} \  
\{a \, M \, b \mid \, a \in M_{n}(\kk),b \in  M_{m}(\kk') \},$$ while the
converse inclusion is true by the previous corollary.
 Actually, for $n=m=2$ and $\kk=\kk'=\cc$, if $M=\pp{1&0\\  0&0}$, 
${\rm Rank}(aMb) \leq1$ for any $a$ and $b$ in $M_n(\cc)$, but for the
fluctuation 
$$f= \{r_1=r_2=1,u_1=v_1=1_2,u_2=v_2=\pp{0&1\\ 1&0} \},\ {\rm
Rank}(\Mf)={\rm Rank}(1_2)=2.$$
\end{rem}

\begin{lem} 
Given a family of $k$ $\rr$-linearly independent matrices $M_i \in
M_{n \times m}(\rr)$, $i=1,\ldots,k$, there exists a fluctuation
$f\in\mathcal{F}_{\rr,\rr}$ such that $\Mf_i=0$ for all $i \neq 1$ and $\Mf_1 \neq
0.$
\end{lem}

\begin{proof}  Let $\{c_i \}_{i\in \{1, \cdots ,p\}}$ be the canonical basis
 of column vectors in
$\rr^p$. (We use abusively the same notation for different $p$s.) Remark 
first that the fluctuation defined for given $r_1,r_2 \in \rr$,
 $i,kÊ\in \{1,\ldots, n\}$ and $j,l \in \{1, \ldots, m \}$ by
\bb
\Mf&:=&r_1Ê\, M +r_2 \; c_ic_k^T \, M \, c_lc_j^T \cr
&=&r_1 \, M + r_2 \, M_{kl} \, c_ic_j^T
\eee
satisfies $(\Mf)_{pq} = r_1 \, M_{pq}$ for all $p \neq i$ and $q \neq j$ and
$(\Mf)_{ij} = r_1 \, M_{ij} + r_2 \, M_{kl}$.

For any $M \in M_{n \times m}(\rr)$ let
\bb
W(M):=\pp{M_{11}\\ \vdots \\ M_{1m} \\M_{21} \\ \vdots \\M_{nm}} \in \rr^{nm}.
\eee
Given a family of matrices 
$M_1,\ldots,M_k \in M_{n \times m}(\rr)$, let $N$ be the
matrix in $M_{nm \times k}(\rr)$ defined by the columns $W(M_i)$:
\bb
N:=(W(M_1) \mid W(M_2) \mid  \ldots \mid  W(M_k)).
\eee
Thus, if a fluctuation $f\in\mathcal{F}_{\rr,\rr}$ is 
defined simultaneously on all
$M_i$'s, $N$ is transformed in 
$f(N):=(W(\Mf_1) \mid W(\Mf_2) \mid \cdots \mid  W(\Mf_k) )$.
By the previous remark, adding a multiple of a line of 
$N$ to a multiple of
another one correspond precisely to a fluctuation. 
Using Gau\ss' method, if the $M_i$'s
are linearly independent (thus $k \leq nm$), so are 
the $W(M_i)$'s and there exists
a fluctuation $f$ such that 
\bb
f(N)=\pp{1_{kÊ\times k} \\ 0_{(nm-k) \times k}}
\eee
since the rank of $N$ is $k$. This means that 
a second fluctuation given by
$\Mg:=c_1c_1^T \, M \, c_1 c_1^T$ will give 
$\Mgf_1=\Mf_1 \neq 0$ while $\Mgf_i=0$
for all $i\neq1$.
\end{proof}

\begin{rem} This lemma is false when the matrices have complex entries: 
Let
$M_1=\pp{i\\1}$ and $M_2=\pp{1\\0}$. Then $M_1$ and
$M_2$ are $\rr$- and $\cc$-linearly independent.
Nevertheless, 
$\Mf_1=0$ always yields $\Mf_2={\rm Im} (\Mf_1)=0$. 
There only remains the following
\end{rem}

\begin{lem} 
Let $M_i \in M_{n \times m}(\cc)$, $i=1,\ldots,k$ be k matrices 
such that their
real and imaginary parts are $2k$ $\rr$-linearly   independent 
matrices. Then there exists a fluctuation
$f\in\mathcal{F}_{\rr,\rr}$ such that $\Mf_1 \neq
0$ and $\Mf_i=0$  for all $i \neq 1$.
\end{lem}

\begin{proof}
According to the previous lemma, there exists a fluctuation
$f\in\mathcal{F}_{\rr,\rr}$ such that ${\rm Re}(\Mf_1) \neq 0$ while 
${\rm Im}(\Mf_1)={\rm Re}(\Mf_i)={\rm Im}(\Mf_i)=0$ for all $i \neq 1$, yielding
the conclusion since real and imaginary extractions 
commute with fluctuations.
\end{proof}

\hfill\break
\indent Within fluctuations, there are the symmetric ones in the following
sense:

\begin{defn}  Let ${\mathcal{F}_{\cc,\cc}}^T:=\{  (r_j,u_j)_{j \in J} \in \rr
\times  U(n)  \mid J {\rm \ finite}\} $ and define fluctuations 
$f^T \in {\mathcal{F}_{\cc,\cc}}^T$ on $n \times n$ square matrices by
$$\MfT:=\sum_j r_j \, u_j{Mu_j}^T.$$
\end{defn}

\begin{lem} Let $M_1,M_2$ be two skewsymmetric matrices in $M_n(\cc)$.
Then 

 i) If the constraints $\MfT_1=0$ for $f \in {{\mathcal{F}}_{\cc,\cc}}^T$ 
always implies $\MfT_2=0$, then $M_2$ is $\cc$-colinear to $M_1$. 

 ii) If $M_2$ is not colinear to $M_1$, then there exists a  fluctuation $f^T \in
{{\mathcal{F}}_{\cc,\cc}}^T$ such that
$\MfT_1=0$ and $\MfT_2\neq 0.$
\end{lem}

\begin{proof} Note that ii) is a consequence of i).\\ To prove i), we may
assume that $M_1$ has the form as in (\ref{antisym}).

\noindent $s:=\pp{0&1\\1&0}$ and
$t:=\pp{1&0\\0&-1}$ are two unitaries satisfying
$wxw^T=-x$ for $w=s,t.$ Define $u=v\oplus1\oplus \cdots\oplus1 \in
U(n)$ with $v=\bigoplus_{i=1}^p v_i \in U(2p)$ where
$v_i \in
\{s,t\}$. Then $M_1+uM_1u^T=0$, so if $M_2$ has the form $\pp{A&B\\ C&D}$,
with A $\in M_{2p}(\cc), B \in M_{2p,n-2p}(\cc), C \in M_{n-2p,2p}(\cc), D
\in M_{n-2p,n-2p}(\cc)$, then $0= M_2+uM_2u^T$. We deduce
$0=A+vAv^T=B+vB=C+Cv^T=D+D$.  Thus choosing $v_s=\bigoplus_{i=1}^p s$
and $v_t=\bigoplus_{i=1}^p t$, we have
$(1_{2p}+v_s)B+(1_{2p}+v_t)B=0$, so $B=0$ since $2\ 1_{2p} + v_s+v_t$ is
invertible. Similarly, $C=D=0$.\\ If $A_{k l}$ is the partition of $A$ in $2
\times 2$ matrices, the constraint $A+vAv^T=0$ implies
$0=A_{kl}+u_kA_{kl} u_l^T$. When $k\neq l$,
$A_{kl}$ is necessarily zero since we may choose independently $u_k$ and
$u_l$ in $\lbrace s,t\rbrace$. When $k=l$,
$A_{kk}=\pp{\alpha_k&\gamma_k\\-\gamma_k&\beta_k}$ where the
$\alpha,\ \beta,\ \gamma$'s are complex numbers, since A is
skewsymmetric. If
$u_k=t$, $0=A_{kk}+u_kA_{kk}u_k=2\pp{\alpha_k&0\\ 0&\beta_k}$ and
$A_{kk}=\gamma_k \, x$. Thus $M_2=\bigoplus_{k=1}^p \gamma_k x
\oplus 0\oplus \cdots \oplus0$ and it remains to prove that $\gamma_k=c\
m_k$ for some constant $c$.\\  Define
\bb u_k&:=&s\oplus1_2\oplus \cdots \oplus1_2\oplus1\oplus \cdots\oplus1,
\cr v_k&:=&1_2\oplus \cdots\oplus s
\oplus \cdots\oplus1_2\oplus1\oplus \cdots \oplus1, \nonumber\\[2mm]
w_k&:=&\pp{0&0&\cdots&0&1_2 \\0&1_2& & &0\\ \vdots& & \ddots &
&\vdots\\0& &&1_2&0 \\ 1_2&0&\cdots&0&0}\oplus1_2\oplus
\cdots\oplus1_2\oplus1
\oplus \cdots\oplus1,
\eee three unitaries where the perturbation in $v_k$ is put at the k-th
entry. Then
$$0=(2m_1)^{-1} \, (M_1-uM_1u^T)-(2m_k)^{-1} \, w_k(M_1-v_kM_1v_k^T)w_k^T$$
and the same relation for
$M_2$ yields
$0=2(\gamma_1 {m_1}^{-1}-\gamma_k {m_k}^{-1})$, so
$\gamma_k=\gamma_1 {m_1}^{-1} \, m_k$ and
$M_2=(\gamma_1 {m_1}^{-1})\,M_1$.
\end{proof}

\begin{tabular}{cccc}
\rxyz{0.7}{
,(10,-5);(5,-5)**\dir{-}?(.6)*\dir{>}
,(15,-10);(10,-10)**\dir{-}?(.6)*\dir{>}
}   &
 $\;$$\;$ \rxyz{0.7}{
,(10,-5);(5,-5)**\dir{-}?(.6)*\dir{>}
,(15,-10);(10,-10)**\dir{-}?(.4)*\dir{<}
} &
 $\;$$\;$ \rxyz{0.7}{
,(10,-5);(5,-5)**\dir{-}?(.6)*\dir{>}
,(10,-15);(15,-15)**\dir{-}?(.6)*\dir{>}
}
 & $\;$$\;$
\rxyz{0.7}{
,(10,-5);(5,-5)**\dir{-}?(.6)*\dir{>}
,(5,-10);(15,-10)**\crv{(10,-13)}?(.6)*\dir{>}
}
 \\
\\ diag. 1 &  $\;$$\;$ diag. 2 &  $\;$$\;$ diag. 3 &  $\;$$\;$ diag. 4 \\
\rxyz{0.7}{
,(10,-5);(5,-5)**\dir{-}?(.6)*\dir{>}
,(10,-5);(10,-15)**\crv{(13,-10)}?(.6)*\dir{>}
}
 &  $\;$$\;$ \rxyz{0.7}{
,(10,-5);(15,-5)**\dir{-}?(.6)*\dir{>}
,(10,-15);(5,-15)**\dir{-}?(.6)*\dir{>}
}
 &  $\;$$\;$ \rxyz{0.7}{
,(5,-10);(5,-15)**\dir{-}?(.6)*\dir{>}
,(10,-15);(5,-15)**\dir{-}?(.6)*\dir{>}
}
 &  $\;$$\;$
\rxyz{0.7}{
,(15,-15);(10,-15)**\dir{-}?(.6)*\dir{>}
,(10,-5);(5,-5)**\dir{-}?(.6)*\dir{>}
,(10,-5);(15,-5)**\dir{-}?(.6)*\dir{>}
}
 \\
\\ diag. 5 &  $\;$$\;$ diag. 6 &  $\;$$\;$ diag. 7 &  $\;$$\;$ diag. 8 \\
\rxyz{0.7}{
,(15,-15);(10,-15)**\dir{-}?(.6)*\dir{>}
,(5,-5);(10,-5)**\dir{-}?(.6)*\dir{>}
,(5,-10);(5,-15)**\dir{-}?(.6)*\dir{>}
}
 &  $\;$$\;$ \rxyz{0.7}{
,(10,-5)*\cir(0.4,0){}*\frm{*}
,(10,-5);(5,-5)**\dir2{-}?(.6)*\dir2{>}
,(15,-15);(10,-15)**\dir{-}?(.6)*\dir{>}
}
 &  $\;$$\;$\rxyz{0.7}{
,(5,-5)*\cir(0.4,0){}*\frm{*}
,(10,-5);(5,-5)**\dir2{-}?(.6)*\dir2{>}
,(15,-15);(10,-15)**\dir{-}?(.6)*\dir{>}
} & 
$\;$$\;$
\rxyz{0.7}{
,(10,-5);(5,-5)**\dir{-}?(.6)*\dir{>}
,(5,-10);(5,-5)**\dir{-}?(.6)*\dir{>}
,(15,-15);(10,-15)**\dir{-}?(.6)*\dir{>}
} \\
\\ diag. 9 &  $\;$$\;$ diag. 10 &  $\;$$\;$ diag. 11 &  $\;$$\;$ diag. 12 \\
\\
\\ &&Fig. 6.1&
\end{tabular}

\begin{tabular}{cccc}
\rxyz{0.7}{
,(10,-5)*\cir(0.4,0){}*\frm{*}
,(10,-5);(5,-5)**\dir2{-}?(.6)*\dir2{>}
,(10,-15);(5,-15)**\dir{-}?(.6)*\dir{>}
}
 & $\;$$\;$ \rxyz{0.7}{
,(5,-5)*\cir(0.4,0){}*\frm{*}
,(10,-5);(5,-5)**\dir2{-}?(.6)*\dir2{>}
,(10,-15);(5,-15)**\dir{-}?(.6)*\dir{>}
}
 &$\;$$\;$ \rxyz{0.7}{
,(10,-5);(5,-5)**\dir{-}?(.6)*\dir{>}
,(5,-10);(5,-5)**\dir{-}?(.6)*\dir{>}
,(10,-15);(5,-15)**\dir{-}?(.6)*\dir{>}
}
 &
$\;$$\;$\rxyz{0.70}{
,(5,-15)*\cir(0.4,0){}*\frm{*}
,(10,-5);(5,-5)**\dir{-}?(.6)*\dir{>}
,(10,-15);(5,-15)**\dir2{-}?(.6)*\dir2{>}
}
\\
\\ diag. 13 &$\;$$\;$ diag. 14 & $\;$$\;$ diag. 15 & $\;$$\;$ diag. 16 \\
\rxyz{0.7}{
,(10,-15)*\cir(0.4,0){}*\frm{*}
,(10,-5);(5,-5)**\dir{-}?(.6)*\dir{>}
,(10,-15);(5,-15)**\dir2{-}?(.6)*\dir2{>}
}
 & $\;$$\;$ \rxyz{0.7}{
,(10,-5)*\cir(0.4,0){}*\frm{*}
,(10,-5);(5,-5)**\dir2{-}?(.6)*\dir2{>}
,(5,-15);(10,-15)**\dir{-}?(.6)*\dir{>}
}
 & $\;$$\;$ \rxyz{0.7}{
,(5,-5)*\cir(0.4,0){}*\frm{*}
,(10,-5);(5,-5)**\dir2{-}?(.6)*\dir2{>}
,(5,-15);(10,-15)**\dir{-}?(.6)*\dir{>}
}
 &
$\;$$\;$ \rxyz{0.7}{
,(10,-5);(5,-5)**\dir{-}?(.6)*\dir{>}
,(5,-10);(5,-5)**\dir{-}?(.6)*\dir{>}
,(5,-15);(10,-15)**\dir{-}?(.6)*\dir{>}
}

\\
\\ diag. 17 & $\;$$\;$ diag. 18 & $\;$$\;$ diag. 19 & $\;$$\;$ diag. 20 \\
\rxyz{0.7}{
,(5,-15)*\cir(0.4,0){}*\frm{*}
,(10,-5);(5,-5)**\dir{-}?(.6)*\dir{>}
,(5,-15);(10,-15)**\dir2{-}?(.6)*\dir2{>}
}
 & $\;$$\;$ \rxyz{0.7}{
,(10,-15)*\cir(0.4,0){}*\frm{*}
,(10,-5);(5,-5)**\dir{-}?(.6)*\dir{>}
,(5,-15);(10,-15)**\dir2{-}?(.6)*\dir2{>}
}
 & $\;$$\;$ \rxyz{0.7}{
,(10,-5);(5,-5)**\dir{-}?(.6)*\dir{>}
,(10,-5);(10,-10)**\dir{-}?(.6)*\dir{>}
,(10,-15);(5,-15)**\dir{-}?(.6)*\dir{>}
}
 &
$\;$$\;$ \rxyz{0.7}{
,(10,-5);(5,-5)**\dir{-}?(.6)*\dir{>}
,(10,-5);(10,-10)**\dir{-}?(.6)*\dir{>}
,(5,-15);(10,-15)**\dir{-}?(.6)*\dir{>}
}
 \\
\\ diag. 21 & $\;$$\;$ diag. 22 & $\;$$\;$ diag. 23 & $\;$$\;$  diag. 24
\\
\\ &&Fig. 6.2&
\end{tabular}

\begin{tabular}{cc}
\rxyd{0.7}{ ,(15,-5);(5,-5)**\crv{(10,-8)}?(.6)*\dir{>}
,(25,-20);(5,-20)**\crv{(15,-24)}?(.6)*\dir{>} } &
$\;$$\;$$\;$$\;$$\;$
\rxyd{0.7}{ ,(15,-5);(5,-5)**\crv{(10,-8)}?(.6)*\dir{>}
,(25,-25);(15,-25)**\crv{(20,-22)}?(.6)*\dir{>}
,(5,-25);(5,-20)**\dir{-}?(.6)*\dir{>} }
\\
\\ diag. 25 & $\;$$\;$$\;$$\;$$\;$ diag. 26
\\
\\ &Fig. 7
\end{tabular}

\begin{center}
\begin{tabular}{cccc}
\rxyz{0.7}{
,(10,-5);(15,-5)**\dir{-}?(.6)*\dir{>}?(1)*\dir{(}
,(5,-15);(10,-15)**\dir{-}?(.6)*\dir{>}
}
   &  & & \\
\\ diag. 27 &  & &  \\
\rxyz{0.7}{
,(10,-5);(5,-5)**\dir{-}?(.6)*\dir{>}?(1)*\dir{(}
,(15,-15);(10,-15)**\dir{-}?(.6)*\dir{>}
}
 & \rxyz{0.7}{
,(10,-5);(5,-5)**\dir{-}?(.6)*\dir{>}?(0)*\dir{)}
,(15,-15);(10,-15)**\dir{-}?(.6)*\dir{>}
}
 && \\
\\ diag. 28 & diag 29 && \\
\rxyz{0.7}{
,(10,-5);(5,-5)**\dir{-}?(.6)*\dir{>}?(1)*\dir{(}
,(10,-15);(5,-15)**\dir{-}?(.6)*\dir{>}
}
 & \rxyz{0.7}{
,(10,-5);(5,-5)**\dir{-}?(.6)*\dir{>}?(0)*\dir{)}
,(10,-15);(5,-15)**\dir{-}?(.6)*\dir{>}
}
 & \rxyz{0.7}{
,(10,-5);(5,-5)**\dir{-}?(.6)*\dir{>}
,(10,-15);(5,-15)**\dir{-}?(.6)*\dir{>}?(1)*\dir{(}
}
 & \rxyz{0.7}{
,(10,-5);(5,-5)**\dir{-}?(.6)*\dir{>}
,(10,-15);(5,-15)**\dir{-}?(.6)*\dir{>}?(0)*\dir{||}
}
 \\
\\ diag. 30 & diag. 31 & diag. 32 & diag. 33 \\
\rxyz{0.7}{
,(10,-5);(5,-5)**\dir{-}?(.6)*\dir{>}?(1)*\dir{(}
,(5,-15);(10,-15)**\dir{-}?(.6)*\dir{>}
}
 & \rxyz{0.7}{
,(10,-5);(5,-5)**\dir{-}?(.6)*\dir{>}?(0)*\dir{)}
,(5,-15);(10,-15)**\dir{-}?(.6)*\dir{>}
}
 &\rxyz{0.7}{
,(10,-5);(5,-5)**\dir{-}?(.6)*\dir{>}
,(5,-15);(10,-15)**\dir{-}?(.6)*\dir{>}?(0)*\dir{)}
}
 & \rxyz{0.7}{
,(10,-5);(5,-5)**\dir{-}?(.6)*\dir{>}
,(5,-15);(10,-15)**\dir{-}?(.6)*\dir{>}?(1)*\dir{||}
}
 \\
\\ diag. 34 & diag. 35 & diag. 36 & diag. 37 \\
\\
\\ && \hspace{-4cm} Fig. 8&
\end{tabular}
\end{center}

\begin{center}
\begin{tabular}{cccc}
\rxy{
,(0,0)*\cir(0.7,0){}
,(5,0)*\cir(0.7,0){}
,(7,0)*{c}
,(0,-4)*{a}
,(5,-4)*{b}
,(10,-2)*{:}
,(0,0);(5,0)**\dir{-}?(.6)*\dir{>}?(1)*\dir{(}
}
&\;\;\;\;
\rxy{
,(0,0)*\cir(0.7,0){}
,(5,0)*\cir(0.7,0){}
,(10,0)*\cir(0.7,0){}
,(0,-5)*\cir(0.7,0){}
,(5,-5)*\cir(0.7,0){}
,(10,-5)*\cir(0.7,0){}
,(12,0)*{c}
,(12,-5)*{\bar{c}}
,(0,-8)*{a}
,(5,-8)*{b}
,(10,-7.9)*{\bar{b}}
,(0,0);(5,0)**\dir{-}?(.6)*\dir{>}
,(10,0);(5,0)**\dir{-}?(.6)*\dir{>}
}
&\;\;\;\;
\rxy{
,(0,0)*\cir(0.7,0){}
,(5,0)*\cir(0.7,0){}
,(10,0)*\cir(0.7,0){}
,(0,-5)*\cir(0.7,0){}
,(5,-5)*\cir(0.7,0){}
,(10,-5)*\cir(0.7,0){}
,(12,0)*{c}
,(12,-5)*{\bar{c}}
,(0,-8)*{a}
,(5,-8)*{b}
,(10,-7.9)*{\bar{b}}
,(0,0);(5,0)**\dir{-}?(.6)*\dir{>}
,(5,-5);(5,0)**\dir{-}?(.6)*\dir{>}
}
& \\ \\ 
\rxy{
,(0,0)*\cir(0.7,0){}
,(5,0)*\cir(0.7,0){}
,(7,0)*{c}
,(0,-4)*{a}
,(5,-4)*{b}
,(10,-2)*{:}
,(5,0);(0,0)**\dir{-}?(.6)*\dir{>}?(0)*\dir{)}
}
&\;\;\;\;
\rxy{
,(0,0)*\cir(0.7,0){}
,(5,0)*\cir(0.7,0){}
,(10,0)*\cir(0.7,0){}
,(0,-5)*\cir(0.7,0){}
,(5,-5)*\cir(0.7,0){}
,(10,-5)*\cir(0.7,0){}
,(12,0)*{c}
,(12,-5)*{\bar{c}}
,(0,-8)*{a}
,(5,-8)*{b}
,(10,-7.9)*{\bar{b}}
,(5,0);(0,0)**\dir{-}?(.6)*\dir{>}
,(5,0);(10,0)**\dir{-}?(.6)*\dir{>}
}
&\;\;\;\;
\rxy{
,(0,0)*\cir(0.7,0){}
,(5,0)*\cir(0.7,0){}
,(10,0)*\cir(0.7,0){}
,(0,-5)*\cir(0.7,0){}
,(5,-5)*\cir(0.7,0){}
,(10,-5)*\cir(0.7,0){}
,(12,0)*{c}
,(12,-5)*{\bar{c}}
,(0,-8)*{a}
,(5,-8)*{b}
,(10,-7.9)*{\bar{b}}
,(5,0);(0,0)**\dir{-}?(.6)*\dir{>}
,(5,0);(5,-5)**\dir{-}?(.6)*\dir{>}
}
& \\ \\
\rxy{
,(0,0)*\cir(0.7,0){}
,(5,0)*\cir(0.7,0){}
,(7,0)*{c}
,(0,-4)*{a}
,(5,-4)*{b}
,(10,-2)*{:}
,(5,0);(0,0)**\dir{-}?(.6)*\dir{>}?(0)*\dir{||}
}
&\;\;\;\;
\rxy{
,(0,0)*\cir(0.7,0){}
,(5,0)*\cir(0.7,0){}
,(10,0)*\cir(0.7,0){}
,(0,-5)*\cir(0.7,0){}
,(5,-5)*\cir(0.7,0){}
,(10,-5)*\cir(0.7,0){}
,(12,0)*{c}
,(12,-5)*{\bar{c}}
,(0,-8)*{a}
,(5,-8)*{b}
,(10,-7.9)*{\bar{b}}
,(5,0);(0,0)**\dir{-}?(.6)*\dir{>}
,(5.3,0);(10,0)**\dir3{-}?(.6)*\dir3{>}
,(5,0);(10,0)**\dir{-}
,(10,0)*\cir(0.4,0){}*\frm{*}
}
&\;\;\;\;
\rxy{
,(0,0)*\cir(0.7,0){}
,(5,0)*\cir(0.7,0){}
,(10,0)*\cir(0.7,0){}
,(0,-5)*\cir(0.7,0){}
,(5,-5)*\cir(0.7,0){}
,(10,-5)*\cir(0.7,0){}
,(12,0)*{c}
,(12,-5)*{\bar{c}}
,(0,-8)*{a}
,(5,-8)*{b}
,(10,-7.9)*{\bar{b}}
,(5,0);(0,0)**\dir{-}?(.6)*\dir{>}
,(5.3,0.15);(10,0.15)**\dir2{-}?(.6)*\dir2{>}
,(5,0);(10,0)**\dir{-}
,(10,-5);(10,0)**\dir{-}?(.6)*\dir{>}
,(10,0)*\cir(0.4,0){}*\frm{*}
}
&\;\;\;\;
\rxy{
,(0,0)*\cir(0.7,0){}
,(5,0)*\cir(0.7,0){}
,(10,0)*\cir(0.7,0){}
,(0,-5)*\cir(0.7,0){}
,(5,-5)*\cir(0.7,0){}
,(10,-5)*\cir(0.7,0){}
,(12,0)*{c}
,(12,-5)*{\bar{c}}
,(0,-8)*{a}
,(5,-8)*{b}
,(10,-7.9)*{\bar{b}}
,(5,0);(0,0)**\dir{-}?(.6)*\dir{>}
,(5,0);(10,0)**\dir{-}?(.6)*\dir{>}
,(10,-5);(10,0)**\dir2{-}?(.6)*\dir2{>}
,(10,0)*\cir(0.4,0){}*\frm{*}
}
\\ \\
&\;\;\;\;
\rxy{
,(0,0)*\cir(0.7,0){}
,(5,0)*\cir(0.7,0){}
,(10,0)*\cir(0.7,0){}
,(0,-5)*\cir(0.7,0){}
,(5,-5)*\cir(0.7,0){}
,(10,-5)*\cir(0.7,0){}
,(12,0)*{c}
,(12,-5)*{\bar{c}}
,(0,-8)*{a}
,(5,-8)*{b}
,(10,-7.9)*{\bar{b}}
,(5,0);(0,0)**\dir{-}?(.6)*\dir{>}
,(5,0);(5,-5)**\dir{-}
,(5,-0.3);(5,-5)**\dir3{-}?(.6)*\dir3{>}
,(5,-5)*\cir(0.4,0){}*\frm{*}
}
&\;\;\;\;
\rxy{
,(0,0)*\cir(0.7,0){}
,(5,0)*\cir(0.7,0){}
,(10,0)*\cir(0.7,0){}
,(0,-5)*\cir(0.7,0){}
,(5,-5)*\cir(0.7,0){}
,(10,-5)*\cir(0.7,0){}
,(12,0)*{c}
,(12,-5)*{\bar{c}}
,(0,-8)*{a}
,(5,-8)*{b}
,(10,-7.9)*{\bar{b}}
,(5,0);(0,0)**\dir{-}?(.6)*\dir{>}
,(10,-5);(5,-5)**\dir{-}?(.6)*\dir{>}
,(5.15,-0.3);(5.15,-5)**\dir2{-}?(.6)*\dir2{>}
,(5,-0);(5,-5)**\dir{-}
,(5,-5)*\cir(0.4,0){}*\frm{*}
}
&\;\;\;\;
\rxy{
,(0,0)*\cir(0.7,0){}
,(5,0)*\cir(0.7,0){}
,(10,0)*\cir(0.7,0){}
,(0,-5)*\cir(0.7,0){}
,(5,-5)*\cir(0.7,0){}
,(10,-5)*\cir(0.7,0){}
,(12,0)*{c}
,(12,-5)*{\bar{c}}
,(0,-8)*{a}
,(5,-8)*{b}
,(10,-7.9)*{\bar{b}}
,(5,0);(0,0)**\dir{-}?(.6)*\dir{>}
,(10,-5);(5,-5)**\dir2{-}?(.6)*\dir2{>}
,(5,0);(5,-5)**\dir{-}?(.6)*\dir{>}
,(5,-5)*\cir(0.4,0){}*\frm{*}
}
\\ \\
\rxy{
,(0,0)*\cir(0.7,0){}
,(5,0)*\cir(0.7,0){}
,(7,0)*{c}
,(0,-4)*{a}
,(5,-4)*{b}
,(10,-2)*{:}
,(0,0);(5,0)**\dir{-}?(.6)*\dir{>}?(1)*\dir{||}
}
&\;\;\;\;
\rxy{
,(0,0)*\cir(0.7,0){}
,(5,0)*\cir(0.7,0){}
,(10,0)*\cir(0.7,0){}
,(0,-5)*\cir(0.7,0){}
,(5,-5)*\cir(0.7,0){}
,(10,-5)*\cir(0.7,0){}
,(12,0)*{c}
,(12,-5)*{\bar{c}}
,(0,-8)*{a}
,(5,-8)*{b}
,(10,-7.9)*{\bar{b}}
,(0,0);(5,0)**\dir{-}?(.6)*\dir{>}
,(10,0);(5.3,0)**\dir3{-}?(.6)*\dir3{>}
,(10,0);(5,0)**\dir{-}
,(10,0)*\cir(0.4,0){}*\frm{*}
}
&
\rxy{
,(5,-2)*{etc.}
}
& \\ \\
&\hspace{4cm} Fig. 9&&
\end{tabular}
\end{center}

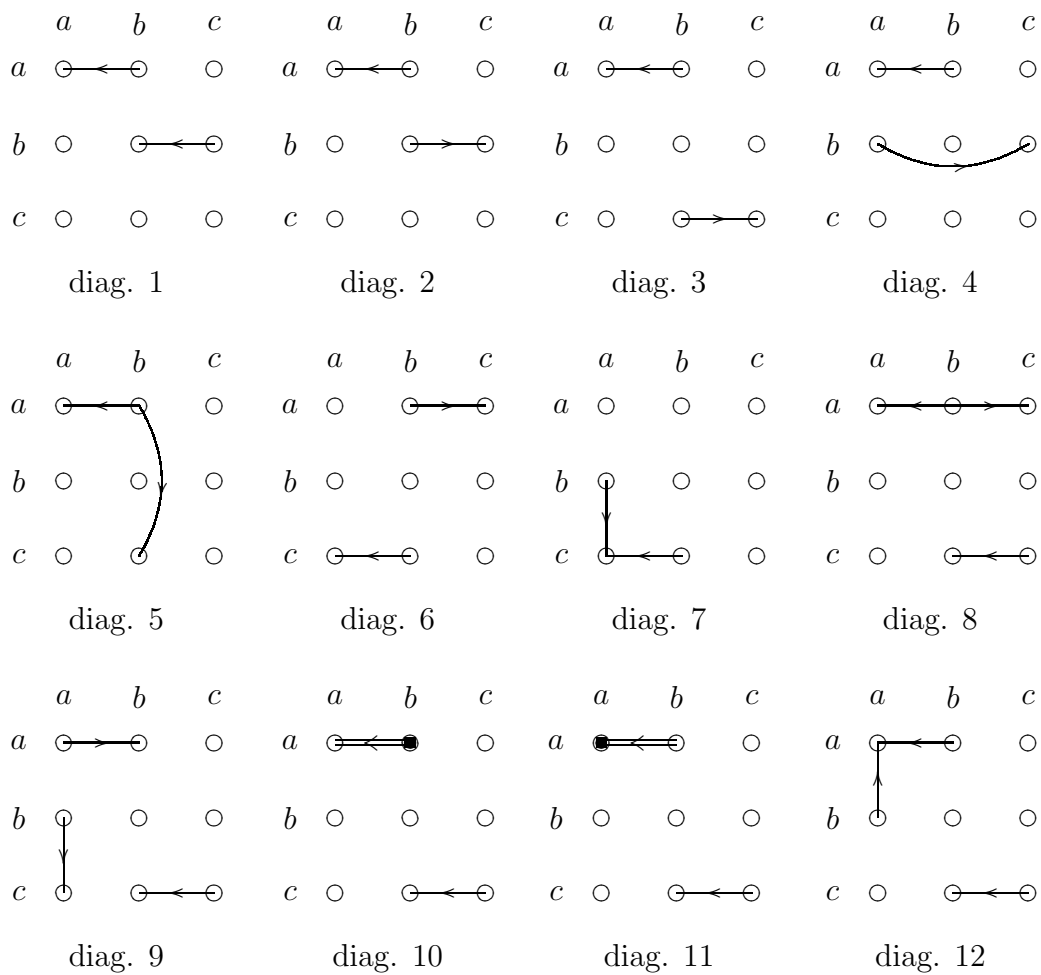
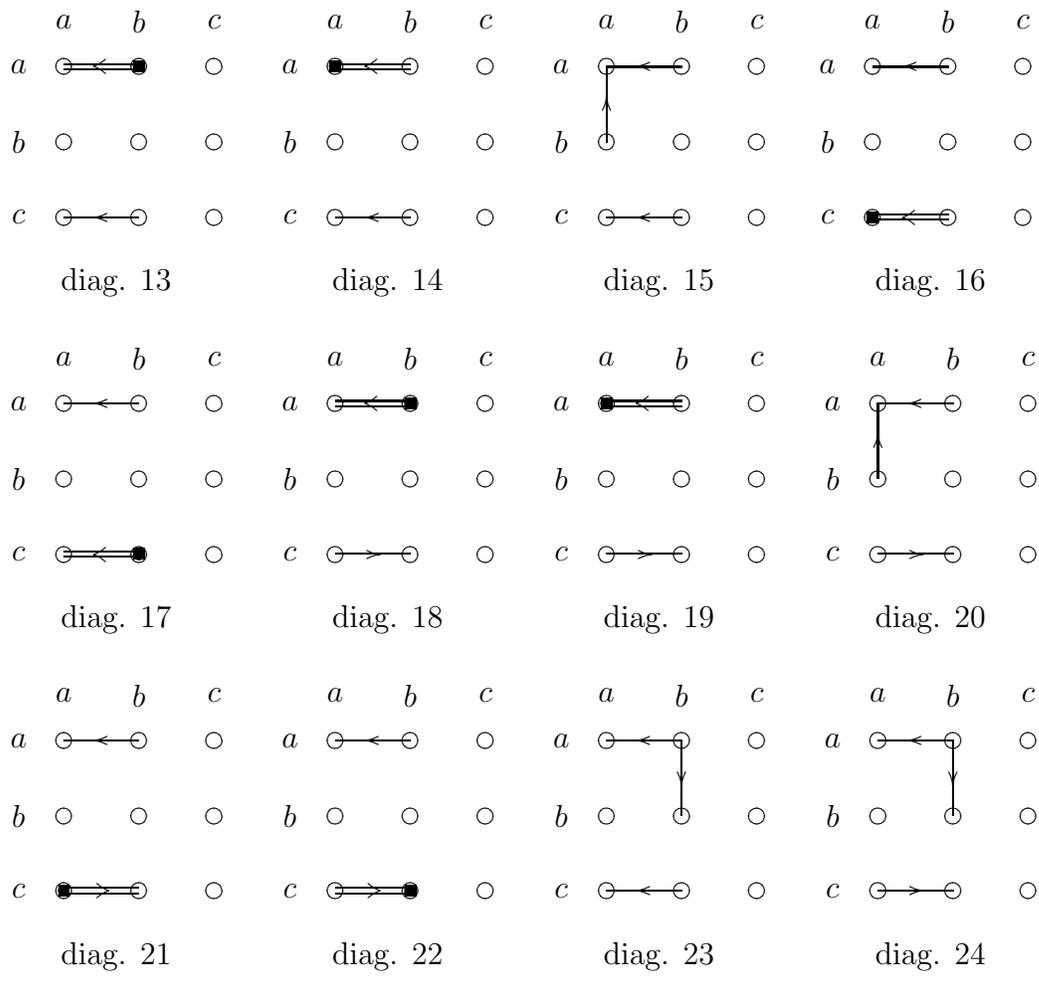
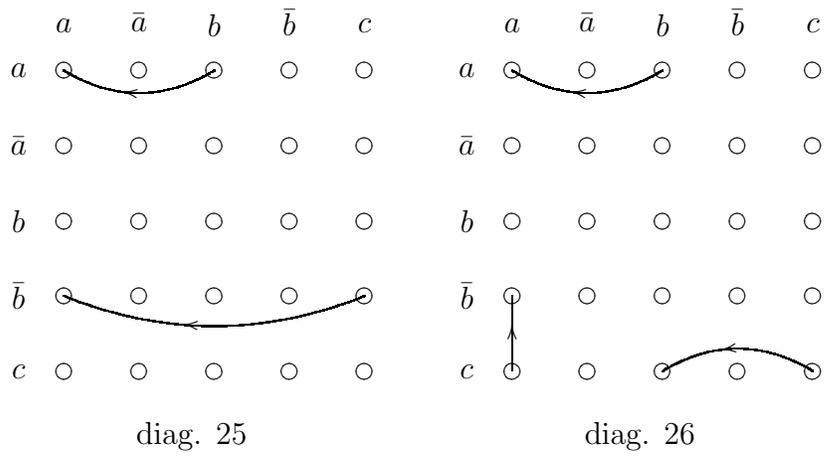
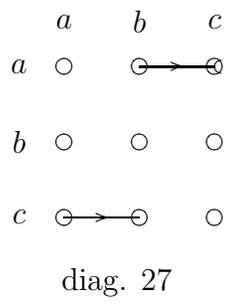
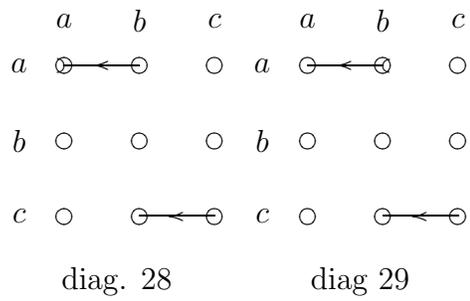
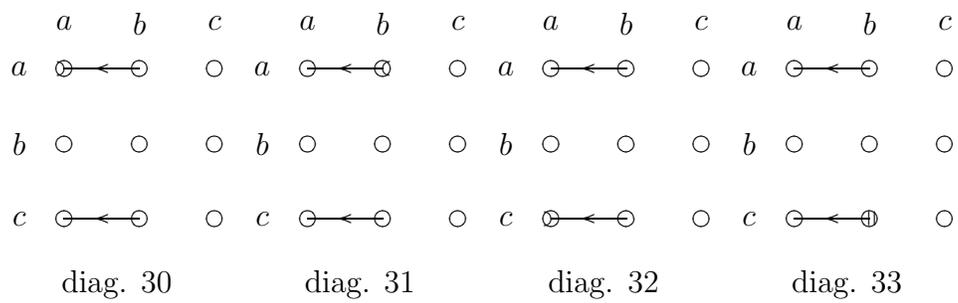
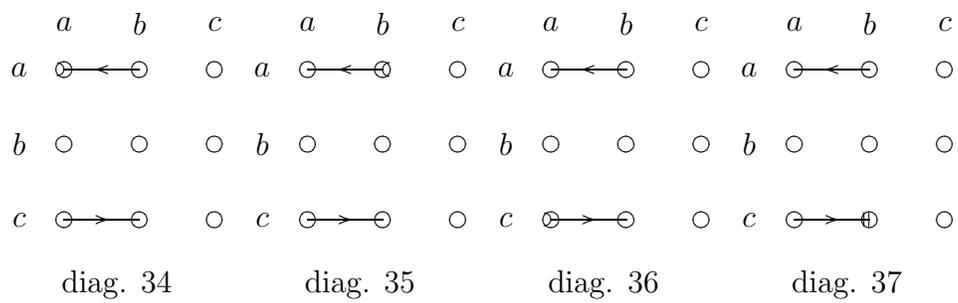
\end{document}